\documentclass[10pt,a4paper]{article}

\usepackage{jheppub}

\usepackage{color}
\usepackage{colordvi}
\usepackage{changes}

\usepackage{epsfig,epsf}
\usepackage{amsmath}
\usepackage{amsthm}
\usepackage{amsfonts}
\usepackage{amssymb}
\usepackage{dsfont}
\usepackage{epstopdf}
\usepackage{multirow}
\usepackage{braket}
\usepackage{tabularx}

\usepackage{rotating}

\usepackage{marvosym}

\usepackage{todonotes}

\usepackage{subcaption}
\usepackage{array}   
\newcolumntype{C}{>{$}c<{$}}

\usepackage{slashed}

\usepackage[active]{srcltx}

\newcommand{\s}[1]{\slashed{#1}}
\newcommand{\m}[1]{\mathbf{#1}}
\newcommand{\w}[1]{\widetilde{#1}}
\newcommand{\f}[1]{\mathcal{#1}}

\newcommand{\Hp}{\text{H}}

\newcommand \widebar [1] {\overline{#1}}

\def\II{\hbox{{1}\kern-.25em\hbox{l}}}

\def\II{\hbox{{1}\kern-.25em\hbox{l}}}

\title{
Two-loop coefficient functions in deeply virtual Compton scattering: flavor-singlet axial-vector and transversity case}

\author[a]{Yao Ji, }
\affiliation[a]{Physik Department T31,\\
James-Franck-Stra\ss e~1, 
Technische Universit\"at M\"unchen,\\
D--85748 Garching, Germany}

\emailAdd{yao.ji@tum.de}

\author[b]{J. Schoenleber}
\affiliation[b]{
   Institut f\"ur Theoretische Physik, Universit\"at
   Regensburg \\ D-93040 Regensburg, Germany}

\emailAdd{jakob.schoenleber@ur.de}

\abstract{
We calculate the two-loop flavor-singlet axial-vector and gluon transversity coefficient functions for deeply virtual Compton scattering in QCD. We observe interesting properties regarding the transcendentality of the transversity coefficient function. 
Our results complete the calculation of the full next-to-next-to-leading order coefficient function in deeply virtual Compton scattering. Numerically, the two-loop corrections in the axial-vector and transversity channel are comparable to their vector counterpart at moderate skewness parameter $\xi$ and hence indispensable for analyzing the upcoming high-precision data from the Electron-Ion Collider. 
}


\begin{document}
\begin{flushright}
{\small
TUM-HEP-1474/23\\
}
\end{flushright}
\maketitle

\section{Introduction}\label{sec:introduction}
The generalized parton distributions (GPDs) \cite{Muller:1994ses, Ji:1996ek, Ji:1996nm, Radyushkin:1997ki} have a rich physical interpretation in terms of transverse spatial probability distribution of partons with a given longitudinal momentum fraction \cite{Burkardt:2002hr}, and are essential for studying the decomposition of the proton spin \cite{Ji:1996ek} and various inter- and multi-parton correlations inside the hadrons. For the case of nucleon (or more generally spin-$\frac{1}{2}$ hardons) targets at leading twist approximation, we can distinguish eight types of GPDs classifiable into three groups:  $H_f, E_f$ (vector), $\w H_f, \w E_f$ (axial-vector), and $H_{f,T}, E_{f,T}, \w H_{f,T}, \w E_{f,T}$ (transversity), where $f = g,u,d,s,...$ runs over parton species. They depend on the averaged parton momentum fraction $x$, the momenta difference in the forward light-like direction between in and outgoing hadron normalized to their total forward momenta, the so-called skewness $\xi$, and the square of the momentum transfer $\Delta_{\perp}$ projected onto the transverse plane. 

The most studied GPD is the vector type $H_f$, whose Fourier transform with respect to $\Delta_{\perp}$ at zero skewness gives the impact parameter distribution of an unpolarized parton in an unpolarized nucleon. However, information on other GPDs is necessary to construct a  complete picture of the inner structure of the nucleons. In particular,
in order to get the impact parameter distribution of a parton of arbitrary polarization in an arbitrarily polarized nucleon, all GPDs are required \cite{Diehl:2005jf}. Another motivation for studying the complete set of GPDs is that their contributions to DVCS can be on the same level for certain observables at leading power, so good control over all channels is required to extract GPDs to a decent accuracy. 

The deeply-virtual Compton scattering (DVCS) \cite{Ji:1996ek, Ji:1996nm}, is considered the ``golden'' channel for experimentally accessing GPDs\footnote{Recently, there have been proposals of computing all flavor-singlet GPDs on the lattice~\cite{Ji:2022thb}.  We refer~\cite{Ji:2013dva, Radyushkin:2017cyf,Ji:2020ect} and references therein for the general theory framework.}. All types of leading-twist GPDs enter in DVCS observables at leading-power besides the quark transversity GPDs $H_{q,T}, E_{q,T}, \w H_{q,T}, \w E_{q,T}$, which are forbidden for spin-1/2 targets by selection rules in the hard scattering. This implies that the transversity (also known as maximal helicity/double flip sector) channel provides a unique window into the transversity gluon distribution inside the nucleons without ``contaminations" from quarks. The GPDs modulate DVCS observables in a convolution with hard coefficient functions (CFs), which can be calculated in perturbation theory as a series in the strong coupling $\alpha_s$. For DVCS five CFs are required at leading-power, denoted by $C_q, C_g, \w C_q, \w C_g, C_{g,T}$. They were first calculated to next-to-leading order (NLO) in \cite{Ji:1997nk, Belitsky:1997rh, Mankiewicz:1997bk, Hoodbhoy:1998vm, Belitsky:2000jk}. The next-to-next-to-leading order (NNLO) correction was calculated for the flavor nonsinglet contribution to $C_q$ in \cite{Braun:2020yib} and to $\w C_q$ in the $\widebar{\rm MS}$-scheme in \cite{Braun:2021grd, Gao:2021iqq}. $C_g$ and the flavor singlet contribution to $C_q$ were calculated in \cite{Braun:2022bpn} soon after in $\widebar{\rm MS}$ as well. In this work, we complete the NNLO studies of DVCS CFs by calculating $\w C_g$, the flavor singlet contribution to $\w C_q$, and the gluon transversity CF $C_{g,T}$ to the two-loop order in QCD. 

This work is organized as follows. In Section \ref{sec: 1} we briefly introduce the kinematic definitions for DVCS. In Section \ref{sec: 2} we discuss the decomposition of DVCS hadronic tensors into the vector, axial-vector and transversity contribution. In Section \ref{sec: 3} we describe the non-trivial details of the matching procedure. The non-triviality comes from the axial-vector sector, where a consistent continuation to $d = 4- 2\epsilon$ dimensions must be made without relying on intrinsically four dimensional objects such as $\gamma_5$ or the Levi-Civita tensor. For the Levi-Civita tensor a natural generalization is to anti-symmetrize the indices that would be contracted with the Levi-Civita tensor in four dimensions. For $\gamma_5$ there is a greater deal of arbitrariness. In this case it is consistent to replace $\gamma^+ \gamma_5$ by an anti-symmetrized combination of three gamma matrices, with one plus and two transverse indices. This corresponds to Larin's scheme \cite{Larin:1993tq}, which can be implemented into the loop calculations in a straightforward fashion. In Section \ref{sec: Larin to MSbar} we discuss how one can convert to another possible scheme for $\gamma_5$, the so-called naive-dimensional regularization (NDR) scheme, where the vector and axial-vector evolution kernels coincide. 
Then in Section \ref{sec: results} we present the main results, the two-loop CFs for the axial-vector and gluon transversity case. Details of the calculation, regarding the additional subtractions to the partonic amplitudes, can be found in Appendix \ref{sec: A}. We briefly discuss the weight reduction of tranversity CF in Appendix \ref{sec: B}.

\section{Preliminaries}
\label{sec: 1}
We consider the DVCS process
\begin{align}
\gamma^*(q) N(p,s) \rightarrow \gamma(q') N(p',s'),
\end{align}
where $\gamma^*(q)$ ($\gamma(q')$) dentotes a virtual (real) photon with momentum $q$ ($q'$). $N(p,s)$ is a hadron state carrying momentum $p$ and helicity $s$. Most commonly $N$ is a proton in experimental studies, but we do not need to require this here, since the CFs do not depend on the target hadron.
We introduce the standard momentum variables
\begin{align}
P = \frac{p+p'}{2}, \qquad \Delta = p' - p
\end{align} 
and Lorentz scalars
\begin{align}
Q^2 = - q^2, \quad x_B = \frac{Q^2}{2p \cdot q}, \quad t = \Delta^2, \quad m^2 = p^2 = p'^2, \quad \xi = \frac{x_B (Q^2 + t)}{2Q^2 - x_B (Q^2 - t)}.
\end{align}
We use light-cone coordinates with respect to two light-like vectors $n, \bar n$ with $n^2 = \bar n^2 = 0$ with normalization condition $n \cdot \bar n = 1$. Without loss of generality we can define
\begin{align}
n^{\mu} = \frac{1}{\sqrt{2}} (1,0,0,1)^{\mu},  \qquad \bar n^{\mu} = \frac{1}{\sqrt{2}} (1,0,0,-1)^{\mu}.
\end{align}
For a generic vector $V$ we define $V^+ = n \cdot V,~ V^- = \bar n \cdot V$ and $V_{\perp}^\mu = V^\mu - V^+ \bar n^\mu - V^- n^\mu$. We will commonly denote a vector $V$ in terms of its light-cone components by
\begin{align}
V = (V^+, V^-, V_{\perp}).
\end{align}

It is useful to choose a preferred frame of reference to analyze the process. 
Our choice of frame in this paper is defined by requirement that $q'^+, q_{\perp}', P_{\perp} = 0$~\cite{Braun:2012bg} which is most suitable for studying leading-power contributions\footnote{For studying higher-power corrections from experimental data, it is more convenient to adopt the reference frame introduced in~\cite{Belitsky:2012ch}. DVCS observables in different frames are related by the so-called kinematic corrections which have been systematically studied to high-accuracy~\cite{Braun:2011zr, Braun:2011dg, Braun:2020zjm, Braun:2022qly} crucial for removing uncertainties due to frame choices~\cite{Guo:2021gru}.}. In particular, $q'$ is taken to be proportional to the light-like vector $n$, whereas the nucleon momenta are proportional to $\bar n$ in the limit of $m^2,t = 0$. Explicitly
\begin{align}
p &= \Big ( (1+\xi) P^+, (1-\xi) \frac{4m^2-t}{8P^+}, \frac{\Delta_{\perp}}{2} \Big ), \notag
\\
p' &= \Big ( (1-\xi) P^+, (1+\xi) \frac{4m^2-t}{8P^+}, - \frac{\Delta_{\perp}}{2} \Big ), \notag
\\
q &= \Big ( -2\xi P^+, \frac{Q^2 + t 
+(4m^2 -t) \xi^2}{4\xi P^+} , - \Delta_{\perp}\Big ),
\\
q' &= \Big ( 0, \frac{Q^2 + t}{4\xi P^+}, 0 \Big ). \notag
\label{eq: momenta}
\end{align}
Here the transverse components of $\Delta_\perp$ are defined only up to rotations in the transverse plane with the constraint that
\begin{align}
\Delta_{\perp}^2 = 4 m^2 \xi^2 + t (1-\xi^2). 
\end{align}
In this work we use the following notation for a perturbative, dimensionally regularized quantity $X(\alpha_s,\epsilon)$
\begin{align}
X = \sum_{n = 0}^{\infty} \Big ( \frac{\alpha_s}{4\pi} \Big )^n X^{(n)} = \sum_{n = 0}^{\infty} \sum_{j = -n}^{\infty} \Big ( \frac{\alpha_s}{4\pi} \Big )^n \epsilon^j X^{(n,j)} .
\end{align}

\section{Decompositon of the hadronic tensor}
\label{sec: 2}
The hadronic part of the DVCS amplitude is given by the hadronic tensor
\begin{align}
T^{\mu \nu} &= i \int d^dx \, e^{-iq \cdot x} \bra{N(p',s')} T \{ j^{\mu}(x) j^{\nu}(0) \} \ket{ N(p,s) },
\label{eq: hadronic tensor}
\end{align}
where $j^{\mu}(x) = \sum_q e_q \bar \psi_q(x) \gamma^{\mu} \psi_q$ with the sum running over active quark flavors $q = u,d,s,...$ of electric charge $e_q$. $T^{\mu \nu}$ is constrained by the electromagnetic Ward identity
\begin{align}
q_{\mu} T^{\mu \nu} = q_{\mu}' T^{\nu \mu} = 0.
\label{eq: WI}
\end{align}
This immediately implies that $T^{\mu +} = 0$ following our convention. Furthermore, the leading twist approximation gives
\begin{align}
q^{\mu} = -2 \xi P^+ \bar n^{\mu} + \frac{Q^2}{4\xi P^+} n^{\mu} + O(t/Q, m^2/Q), \qquad q'^{\mu} = \frac{Q^2}{4\xi P^+} n^{\mu} + O(t/Q,m^2/Q).
\end{align}
Thus from~\eqref{eq: WI}, we obtain,
\begin{align}
T^{+ \nu} = \Big ( - \frac{q^+}{q^-} + O(t/Q,m^2/Q) \Big ) T^{- \nu}  .
\end{align}
It will be convenient to write the projector onto the transverse plane in the following covariant form
\begin{align}
g_{\perp}^{\mu \nu} = g^{\mu \nu} - n^{\mu} \bar n^{\nu} - \bar n^{\mu} n^{\nu}\, .
\end{align}
Using eq. \eqref{eq: WI} gives
\begin{align}
T^{\mu \nu} &= g_{\perp}^{\mu \mu'} g_{\perp}^{\nu \nu'} T_{\mu' \nu'} + \Big ( n^{\mu} - \frac{q^+}{q^-} \bar n^{\mu} \Big ) n^{\nu} T^{--} + \Big ( n^{\mu} - \frac{q^+}{q^-} \bar n^{\mu} \Big ) g_{\perp}^{\nu \nu'} T^-_{~\, \nu'} + n^{\nu} g_{\perp}^{\mu \mu'} T_{\mu'}^{~\, - }
\\
&\qquad + \text{ higher-power}. \nonumber
\end{align}
A few comments are in order. 
Firstly, note that the structures proportional to $n^{\nu}$ do not contribute in DVCS. Indeed, the hadronic tensor will enter the scattering amplitude contracted with the physical polarization vector $\varepsilon_{\nu}^*(q')$ of the outgoing photon, which is in the transverse plane, so that $\varepsilon_{\nu}^*(q') n^{\nu} = 0$.
Secondly, the projection $g_{\perp}^{\nu \nu'} T^-_{~\, \nu'}$ is of subleading power. To see this, note that the $\nu'$ sits at a hard scattering vertex, where the leading twist approximation $m^2 = t = 0$ applies, which implies that all the momenta $p, \, p', \, q, \, q'$ have zero transverse component. The leading term in $T^-_{~\, \nu'}$ on the other hand, having only a single Lorentz index, must therefore be proportional to external momenta with zero transverse component, so that the leading contribution to $g_{\perp}^{\nu \nu'} T^-_{~\, \nu'}$ vanishes.

Hence we conclude
\begin{align}
T^{\mu \nu} &= g_{\perp}^{\mu \mu'} g_{\perp}^{\nu \nu'} T_{\mu' \nu'} + \text{ higher-power}. 
\end{align}
A further decomposition of the tensor product $g_{\perp}^{\mu \mu'} g_{\perp}^{\nu \nu'}$ can be made in terms of the following Lorentz structures
\begin{align}
\tau_{\perp}^{\mu \nu \mu' \nu'} &= \frac{1}{2} g_{\perp}^{\mu \mu'} g_{\perp}^{\nu \nu'} + \frac{1}{2} g_{\perp}^{\mu \nu'} g_{\perp}^{\nu \mu'} - \frac{1}{d-2} g_{\perp}^{\mu \nu} g_{\perp}^{\mu' \nu'},
\\
{} \chi_{\perp}^{\mu \nu \mu' \nu'} &= \frac{1}{2} g_{\perp}^{\mu \mu'} g_{\perp}^{\nu \nu'} - \frac{1}{2} g_{\perp}^{\mu \nu'} g_{\perp}^{\nu \mu'} .
\label{eq: Eperp def}
\end{align}
Then we have trivially
\begin{align}
g_{\perp}^{\mu \mu'} g_{\perp}^{\nu \nu'} = \frac{1}{d-2} g_{\perp}^{\mu \nu} g_{\perp}^{\mu' \nu'}  \, + \, {} \chi_{\perp}^{\mu \nu \mu' \nu'}  + \tau_{\perp}^{\mu \nu \mu' \nu'}.
\label{eq: perp decomp}
\end{align}
Note that the three basis tensor structures in eq. \eqref{eq: perp decomp} are mutually orthogonal and we have
\begin{align}
{} \chi_{\perp}^{\mu \nu \rho \sigma} {} \chi_{\perp, \rho \sigma}^{\quad ~~ \mu' \nu'} = {} \chi_{\perp}^{\mu \nu \mu' \nu'},
\qquad \tau_{\perp}^{\mu \nu \rho \sigma} \tau_{\perp, \rho \sigma}^{\quad ~~ \mu' \nu'} = \tau_{\perp}^{\mu \nu \mu' \nu'}.
\end{align}
It is easy to see that in $d = 4$
\begin{align}
{} \chi_{\perp}^{\mu \nu \mu' \nu'} \stackrel{d=4}{=} \frac{1}{2}\varepsilon_{\perp}^{\mu \nu} \varepsilon_{\perp}^{\mu' \nu'},
\label{eq: E d = 4 id}
\end{align}
where
\begin{align}
\varepsilon_{\perp}^{\mu \nu} = \varepsilon^{\mu \nu \mu' \nu'} \bar  n_{\mu'} n_{\nu'}
\label{eq: epsperp def}
\end{align}
and $\varepsilon^{\mu \nu \mu' \nu'}$ is the four-dimensional Levi-Civita tensor with convention $\varepsilon^{0123} = 1$, so that $\varepsilon_{\perp}^{12} = 1$.
Thus the decomposition eq. \eqref{eq: perp decomp} reduces to the conventional four-dimensional decomposition used in \cite{Belitsky:2000jk}
\begin{align}
g_{\perp}^{\mu \mu'} g_{\perp}^{\nu \nu'} \stackrel{d=4}{=} \frac{1}{2} g_{\perp}^{\mu \nu} g_{\perp}^{\mu' \nu'}  \, + \, \frac{1}{2} \varepsilon_{\perp}^{\mu \nu} \varepsilon_{\perp}^{\mu' \nu'}  + \tau_{\perp}^{\mu \nu \mu' \nu'}.
\label{eq: perp decomp 4d}
\end{align}
In four dimensions we can use the decomposition in eq. (\ref{eq: perp decomp 4d}) to get
\begin{align}
T^{\mu \nu} = - g_{\perp}^{\mu \nu} \f V - i \varepsilon_{\perp}^{\mu \nu} \f A + \f T^{\mu \nu} + \text{ higher-power},
\end{align}
where
\begin{align}
\f V = - \frac{1}{2} g_{\perp}^{\mu \nu} T_{\mu \nu}, \quad \f A = \frac{1}{2} i \varepsilon_{\perp}^{\mu \nu}  T_{\mu \nu}, \quad \f T^{\mu \nu} = \tau_{\perp}^{\mu \nu \mu' \nu'}  T_{\mu' \nu'},
\label{eq: V, A, T}
\end{align}
are the vector, axial-vector and transversity Compton form factors (CFFs) respectively. 

To the leading-power accuracy these contributions can be written in terms of a convolution of CFs with leading-twist GPDs \cite{Radyushkin:1997ki,Collins:1998be, Ji:1998xh},
\begin{align}
\f V(\xi,t) &= \sum_q \int_{-1}^1 \frac{dx}{\xi} \, C_q(x/\xi,Q,\mu_{\rm UV}, \mu_F) F_q(x,\xi,t,\mu_F) \nonumber
\\
&\quad + \int_{-1}^1 \frac{dx}{\xi^2} \, C_g(x/\xi,Q,\mu_{\rm UV}, \mu_F) F_g(x,\xi,t,\mu_F), \nonumber
\\
\f A(\xi,t) &= \sum_q\int_{-1}^1 \frac{dx}{\xi} \, \w C_q(x/\xi,Q,\mu_{\rm UV}, \mu_F) \w F_q(x,\xi,t,\mu_F) 
\label{eq: factorization}
\\
&\quad + \int_{-1}^1 \frac{dx}{\xi^2} \, \w C_g(x/\xi,Q,\mu_{\rm UV}, \mu_F) \w F_g(x,\xi,t,\mu_F),\nonumber
\\
\f T^{\mu \nu}(\xi,t) &= \int_{-1}^1 \frac{dx}{\xi^2} C_{g,T}(x/\xi,Q,\mu_{\rm UV}, \mu_F) F_{g,T}^{\mu \nu}(x,\xi,t,\mu_F). \nonumber
\end{align}
Here $\mu_{\rm UV}$ is the UV renormalization scale that appears in the Lagrangian and enters the scale dependence of $\alpha_s$.  $\mu_F$ is the factorization scale that separates degrees of hard modes of virtuality $\sim Q^2$ from the collinear modes of virtuality $\sim t,m^2$. Note that $\f V, \f A, \f T^{\mu \nu}$ being given by hadronic matrix elements of conserved currents, are finite and scale independent. In practice, however, these contributions \textit{do} have scale dependence due to finite $\alpha_s$ truncation in pQCD calculations. The degree of the scale dependence  therefore measures the uncertainty of the pQCD predictions. Throughout this work, we will set $\mu_{\rm UV} = \mu_F = \mu$ and we will frequently omit the functional dependence on $\mu$ for notational brevity. 

The GPDs can be defined in terms of renormalized hadronic matrix elements of light-ray operators 
\begin{align}
\f O_{q,\alpha \beta}(z^-) &= \bar \psi_{q,\beta}(-z^- n/2) W_{n,F}(-z^- /2,z^- /2) \psi_{q,\alpha}(z^-n/2), \nonumber
\\
\f O_g^{\mu \nu}(z^-) &= G^{\mu +}(-z^-n/2) W_{n,A}(-z^- /2,z^- /2) G^{+ \nu}(z^-n/2),
\end{align}
where $\psi_q$ is a quark field of flavor $q$ (not to be confused with the virtual photon momentum), $G^{\mu\nu}$ is the gluon field strength tensor, and
\begin{align}
W_{n, i}(z_1,z_2) = \text P \exp \Big ( ig \int_{z_2}^{z_1} ds \, A^{+,a}(sn) T_i^a \Big ), \hspace*{10mm} i=F, A,
\end{align}
with $W_{n,F}$ ($W_{n,A}$) being a collinear Wilson line in the fundamental (adjoint) representation ensuring gauge invariance of the nonlocal operators.

The standard definition of the renormalized GPDs in $d=4$ reads
\begin{align}
F_q(x,\xi,t) &= \frac{1}{2} \int \frac{dz^-}{2\pi} e^{ix P^+ z^-} \bra {N(p',s')} 
\text{tr}( \gamma^+ \f O_q(z^-) ) \ket{ N(p,s)}, \nonumber
\\
\w F_q(x,\xi,t) &= \frac{1}{2} \int \frac{dz^-}{2\pi} e^{ix P^+ z^-} \bra {N(p',s')} \text{tr}( \gamma^+ \gamma_5 \f O_q(z^-) ) \ket{ N(p,s)}, \nonumber
\\
F_g(x,\xi,t) &= \frac{1}{P^+} \int \frac{dz^-}{2\pi} e^{ixP^+ z^-} \bra {N(p',s')} g_{\perp}^{\mu' \nu'}  \f O_{g,\mu' \nu'}(z^-) \ket {N(p,s)},  
\label{eq: GPDs def}
\\
\w F_g(x,\xi,t) &= \frac{1}{P^+} \int \frac{dz^-}{2\pi} e^{ixP^+ z^-} \bra {N(p',s')} i\varepsilon_{\perp}^{\mu' \nu'} \f O_{g,\mu' \nu'}(z^-) \ket {N(p,s)},   \nonumber
\\
F_{g,T}^{\mu \nu}(x,\xi,t) &= \frac{1}{P^+} \int \frac{dz^-}{2\pi} e^{ixP^+ z^-}
 \bra {N(p',s')} \tau_{\perp}^{\mu' \nu' \mu \nu} \f O_{g,\mu' \nu'}(z^-) \ket {N(p,s)}.\nonumber
\end{align}
For spin-$\frac{1}{2}$ targets, the GPDs in eq. \eqref{eq: GPDs def} can be further decomposed into the GPDs\\ $H_f,\w H_f, E_f, \w E_f, H_{g,T}, E_{g,T}, \w H_{g,T}, \w E_{g,T}$, mentioned in the introduction, see e.g. \cite{Diehl:2003ny,Belitsky:2005qn} for details.

\section{Dimensional regularization}
\label{sec: 3}

To calculate the CFs we use the so-called ``matching'' procedure, where we evaluate both sides of eq. \eqref{eq: factorization} using on-shell external partonic states. 
There are two types of divergences that appear in the perturbative expansion of the on-shell partonic amplitudes, which are commonly referred to as ``bare'' CFs.
\begin{itemize}
\item UV divergences, which are removed by rewriting the perturbative series in terms of the renormalized coupling. This is sufficient if the external parton propagators are removed according to the Lehmann-Zimmermann-Symanzik formula. In case the renormalized Lagrangian is employed to compute the CFs, the UV divergences would disappear. In our studies, we instead adopt the bare Lagrangian for computational simplicity.
\item IR divergences due to having massless and collinear partons. They are removed from the terms corresponding to the renormalization of the GPD. In the graphical treatment, they correspond to the so-called double-counting subtractions.
\end{itemize}
The divergences are most conveniently regularized using dimensional regularization (dim. reg.) with $d = 4-2\epsilon$ and subtracted using the $\overline{\rm MS}$ prescription. It is also most convenient to treat UV and IR divergences using the same $\epsilon = \epsilon_{\rm UV} = \epsilon_{\rm IR}$. 
The matching procedure is covered in detail in Appendix \ref{sec: A}. 

In order to use dim. reg., we need to choose a $d$-dimensional representation for $T^{\mu \nu}$. As usual, this introduces an ambiguity, which must be treated using a consistent scheme. Firstly, we use the decomposition in eq. \eqref{eq: perp decomp}. Since everything is written only in terms of $g_{\perp}^{\mu \nu}$ it can be naturally extended to $d$ dimensions by augmenting the transverse space from $2$ to $d-2$ dimensions, such that $-g_{\perp}^{\mu \nu}$ is the unit matrix in that space. This gives
\begin{align}
\widehat T^{\mu \nu}_f = - g_{\perp}^{\mu \nu} \widehat{\f V}_f - i \widehat{\f A}_f^{\mu \nu} + \widehat{\f T}_f^{\mu \nu},
\end{align}
where the index $f = g,u,d,s,...$ denotes the parton species of the external state and the hat indicates that we are dealing with dimensionally regularized \textit{partonic} quantities. The partonic vector, axial-vector and transversity contributions are then given by
\begin{align}
\widehat{\f V}_f = - \frac{g_{\perp}^{\mu \nu}}{d-2}  \widehat{T}_{f,\mu \nu}, \quad \widehat{\f A}_f^{\mu \nu} = i {} \chi_{\perp}^{\mu \nu \mu' \nu'} \widehat{T}_{f,\mu' \nu'}, \quad \widehat{\f T}_f^{\mu \nu} = \tau_{\perp}^{\mu \nu \mu' \nu'}  \widehat{T}_{f,\mu' \nu'}.
\label{eq: V, A, T}
\end{align}
The factorization theorem in terms of the \textit{partonic} quantities reads
\begin{align}
\widehat {\f V}_f &= \sum_q \int_{-1}^1 \frac{dx}{\xi} \, C_q(x/\xi,Q) \widehat F_{q/f}(x,\xi) + \int_{-1}^1 \frac{dx}{\xi^2} \, C_g(x/\xi,Q) \widehat F_{g/f}(x,\xi), \nonumber
\\
\widehat {\f A}_f^{\mu \nu} &= \sum_q\int_{-1}^1 \frac{dx}{\xi} \, \w C_q(x/\xi,Q) \widehat {\w F}_{q/f}^{\mu \nu}(x,\xi)+ \int_{-1}^1 \frac{dx}{\xi^2} \, \w C_g(x/\xi,Q) \widehat {\w F}_{g/f}^{\mu \nu}(x,\xi),
\label{eq: factorization partonic}
\\
\widehat {\f T}_f^{\mu \nu} &= \int_{-1}^1 \frac{dx}{\xi^2} C_{g,T}(x/\xi, Q) \widehat F_{g/f,T}^{\mu \nu}(x,\xi) \nonumber
\end{align}
with the same CFs as in eq. \eqref{eq: factorization}, since the hard scattering does not depend on the external state. After replacing $p,p'$ with the leading twist momenta, the partonic GPDs are defined in the same way as in eq. \eqref{eq: GPDs def}, with the exception of the axial-vector sector. We choose 
\begin{align}
\widehat{\w F}_{q/f}^{\mu \nu}(x,\xi) &= \frac{1}{2} \int \frac{dz^-}{2\pi} e^{ix P^+ z^-} \bra {f(p',s')} i{} \chi_{\perp, \mu' \nu'}^{\mu \nu}  \text{tr}( \gamma^+ \gamma^{\mu'} \gamma^{\nu'} \f O_q(z^-) ) \ket{ f(p,s)}, \nonumber
\\
\widehat{\w F}_{g/f}^{\mu \nu} (x,\xi,t) &= \frac{1}{P^+} \int \frac{dz^-}{2\pi} e^{ixP^+ z^-} \bra {f(p',s')} 2 i {} \chi_{\perp}^{\mu \nu \mu' \nu'} \f O_{g,\mu' \nu'}(z^-) \ket {f(p,s)},
\label{eq: axial GPDs partonic}
\end{align}
which are immediately generalizable to generic $d$-dimensions advantageous for loop calculations in dim. reg.
In four dimensions the definitions in eq. \eqref{eq: axial GPDs partonic} coincide with the definitions in eq. \eqref{eq: GPDs def} up to a factor of $\varepsilon_{\perp}^{\mu \nu}$, i.e. they can be recovered by taking $\mu = 1$ and $\nu = 2$. For the gluon case, this is immediately seen by using eq. \eqref{eq: E d = 4 id}. For the quark case, note that
\begin{align}
i {} \chi_{\perp, \mu \nu}^{12} \gamma^+ \gamma^{\mu} \gamma^{\nu} \stackrel{d=4}{=} \gamma^+ \gamma_5. 
\label{eq: L scheme}
\end{align}
This choice to represent $\gamma_5$ in $d$ dimensions corresponds to Larin's scheme \cite{Larin:1993tq}.

We mention that recently there has been some work on DVCS \cite{Bhattacharya:2022xxw, Bhattacharya:2023wvy} that calculates the one-loop CF using a finite $t$ regulator. For the bare CF one finds a simple pole $\frac{1}{t}$ contribution in the vector and axial-vector case, connected to the trace and chiral anomaly respectively. This pole is no contradiction to factorization, since all such IR divergences in the bare CF are removed by double-counting subtractions.
Indeed, the authors find that, after applying these subtractions, all singularities as $t \rightarrow 0$ are removed, so that after the Taylor expansion one obtains a finite renormalized CF. In particular, the renormalized CF does not depend on the regularization procedure \footnote{up to ambiguities concerning $\gamma_5$ and the Levi-Civita tensor, which are a different issue.}. This is of course perfectly in line with our treatment, since dim. reg. already regulates all IR divergences in the bare CF. For a discussion of the double-counting subtractions for DVCS at two-loops we refer to Appendix \ref{sec: A}.


\section{Relation to naive-dimensional regularization scheme}
\label{sec: Larin to MSbar}
Another commonly used choice for $\gamma_5$ is the naive-dimensional regularization (NDR) scheme, often simply called ``$\overline{\rm MS}$-scheme'', in which the vector and axial-vector flavor-nonsinglet evolution kernels coincide. This condition comes naturally if a renormalization scheme other than dim. reg. is used.
It is clear that adopting the NDR scheme will not lead to any algebraic inconsistencies since $\gamma_5$ is absent in the Dirac traces in QCD.

It is possible to convert the CFs between different schemes by enforcing the matrix elements of the so-called evanescent operators to zero. These operators are the differences between operators describing the same physical object but in different schemes. Hence they coincide in $d=4$ but are distinct for $d\neq4$, e.g., the two Dirac structures in eq.~\eqref{eq: L scheme}. The condition that the matrix elements of the evanescent operators must vanish implies (finite) renormalization constants allowing one to translate the CFs from one scheme to another. Let us consider the $\gamma\gamma^*\to \pi^0$ transition form factor as a concrete example\footnote{This process is directly related to the nonsinglet axial-vector contribution~\cite{Braun:2021grd}.}. The evanescent operator necessary to relate the nonsinglet (NS) CFs in Larin and $\widebar{\rm MS}$ scheme takes the form~\cite{Wang:2017ijn, Gao:2021iqq},
\begin{align}
{\cal O}^{\mu\nu}_{{\rm NS}, E}=\bar\psi(-z^-n/2)W_{n,F}(-z^-/2,z^-/2)\gamma^+\left(\frac{[\gamma^{\mu}_{\perp},\gamma^{\nu}_{\perp}]}2-i\varepsilon^{\mu\nu}_\perp\gamma_5\right)\psi(z^-n/2)\label{eq:evanO}
\end{align}
with $\gamma_5$ subscribing to the NDR prescription. The first and second terms in the parentheses correspond to definitions of axial-vector in Larin and $\widebar{\rm MS}$ scheme, respectively.
It is clear that in $d=4$, the operator ${\cal O}_{{\rm NS}, E}^{\mu\nu}$ is identically zero due to simple Dirac algebra. However, in $d=4-2\epsilon$ dimensions, the operator is nonvanishing and proportional to $\epsilon$. Enforcing the partonic matrix element of the IR-finite UV-renormalized evanescent operator to vanish identically in four dimensions to all orders in perturbation theory~\cite{Buras:1992tc,Dugan:1990df},
\begin{align}
\bra{\bar\psi(k_1)}{\cal O}_{{\rm NS},E}^{\mu\nu}\ket{\psi(k_2)} = 0\, ,
\end{align}
allows us to determine its renormalization constant $Z_{{\rm NS},E}$ which is currently known to the ${\cal O}(\alpha_s^2)$ by direct calculation~\cite{Gao:2021iqq}. 
$Z_{{\rm NS},E}$ is then used to translate the nonsinglet axial-vector CF from Larin scheme (L) to the $\widebar{\rm MS}$ scheme. At ${\cal O}(\alpha_s)$, we have,
\begin{align}
    \widetilde C_{q,\rm NS}^{(1),\widebar{\rm MS}}&= \widetilde C_{q,\rm NS}^{(1),\rm L} - C_q^{(0)} \ast Z_{{\rm NS},E}^{(1)}\, ,
\end{align}
with higher-order expressions taking a similar but more complex form. The one-loop renormalization constant $Z_{{\rm NS},E}^{(1)}$ for the evanescent operator ${\cal O}_{{\rm NS},E}^{\mu\nu}$ is finite in $\epsilon$ and $*$ denotes the convolution over partonic momentum fractions.
For more technical details, we refer~\cite{Gao:2021iqq, Wang:2017ijn, Braun:2021tzi} and references therein.

For our current case, to convert our Larin scheme results for both $\widetilde C_q(x/\xi, Q)$ and $\widetilde C_g(x/\xi, Q)$ to the $\widebar{\rm MS}$ scheme require computing the singlet evanescent contribution to the two-loop order. This is beyond the scope of our current paper and therefore we leave it for future studies.

\section{Results for the coefficient functions}
\label{sec: results}
In the following we set $d = 4$.
It will be convenient to introduce the variables
\begin{align}
z = \frac{1}{2} (1- x/\xi), \qquad \bar z = 1-z, \qquad L = \log \frac{\mu^2}{Q^2},\qquad \beta_0 = \frac{11}{3} C_A - \frac43T_F\, n_f
\end{align}
for presenting our main results. In this way, at the tree-level the CFs read
\begin{align}
C_q^{(0)}(x/\xi) &= \frac{e_q^2}{2} \Big ( \frac{1}{z} - \frac{1}{\bar z} \Big ), \quad \w C_q^{(0)}(x/\xi) = \frac{e_q^2}{2} \Big ( \frac{1}{z} + \frac{1}{\bar z} \Big ), \quad C_g^{(0)} = \w C_g^{(0)} = C_{g,T}^{(0)} = 0.
\end{align}
The one-loop contributions are \cite{Ji:1997nk, Belitsky:1997rh, Mankiewicz:1997bk, Hoodbhoy:1998vm, Belitsky:2000jk}
\begin{align}
C_q^{(1)}(x/\xi, L) &= \Big[e_q^2 \frac{C_F }{2z}  \Big ( - L( 3 + 2\log z) + \log^2 z - 3 \frac{z}{\bar z} \log z  - 9 \Big )\Big]  - [z \leftrightarrow \bar z] , \nonumber
\\
\w C_{q}^{(1)}(x/\xi, L) &= \Big[ e_q^2 \frac{ C_F}{2z} \Big ( - L( 3 + 2\log z) + \log^2 z + 7 \frac{z}{\bar z} \log z  - 9 \Big )  \Big ] + [z \leftrightarrow \bar z],\nonumber
\\
C_g^{(1)}(x/\xi, L) &= \Big[\Big (  \sum_q e_q^2 \Big )  \frac{T_F  }{2\bar z z} \Big ( L \frac{z}{\bar z} \log z - \frac{z}{2\bar z} \log^2 z  + \frac{1+z}{\bar z} \log z   \Big)\Big ] + [z \leftrightarrow \bar z] ,
\label{eq: oneloop results}
\\
\w C_{g}^{(1)}(x/\xi, L) &= \Big[\Big (  \sum_q e_q^2 \Big )  \frac{T_F}{2\bar z z} \Big ( - L \frac{z}{\bar z} \log z + \frac{z}{2\bar z} \log^2 z + \frac{1-3z}{\bar z} \log z\Big)  \Big ] - [z \leftrightarrow \bar z]\nonumber
\\
C_{g,T}^{(1)}(x/\xi, L) &= \Big (  \sum_q e_q^2 \Big )  \frac{T_F}{z \bar z}. \nonumber
\end{align}

The calculation procedure is completely analogous to the vector case in \cite{Braun:2022bpn}. For the evaluation of Dirac traces and Lorentz contractions \texttt{FORM} \cite{Vermaseren:2000nd} as well as in-house routines were used. 
The resulting set of scalar integrals was reduced to 
twelve master integrals making use of the integration-by-parts relations, performed with \texttt{FIRE} \cite{Smirnov:2019qkx}.
No new master integrals appeared compared to the vector nonsinglet case.
They were first calculated in \cite{Gao:2021iqq}. For the calculation of the double-counting/IR subtractions, see Appendix \ref{sec: A}, the program \texttt{HyperInt} \cite{Panzer:2014caa} and in-house routines were used for this task.

\subsection*{Two-loop axial-vector CFs}
\begin{figure*}[ht]
~~\includegraphics[scale=0.245]{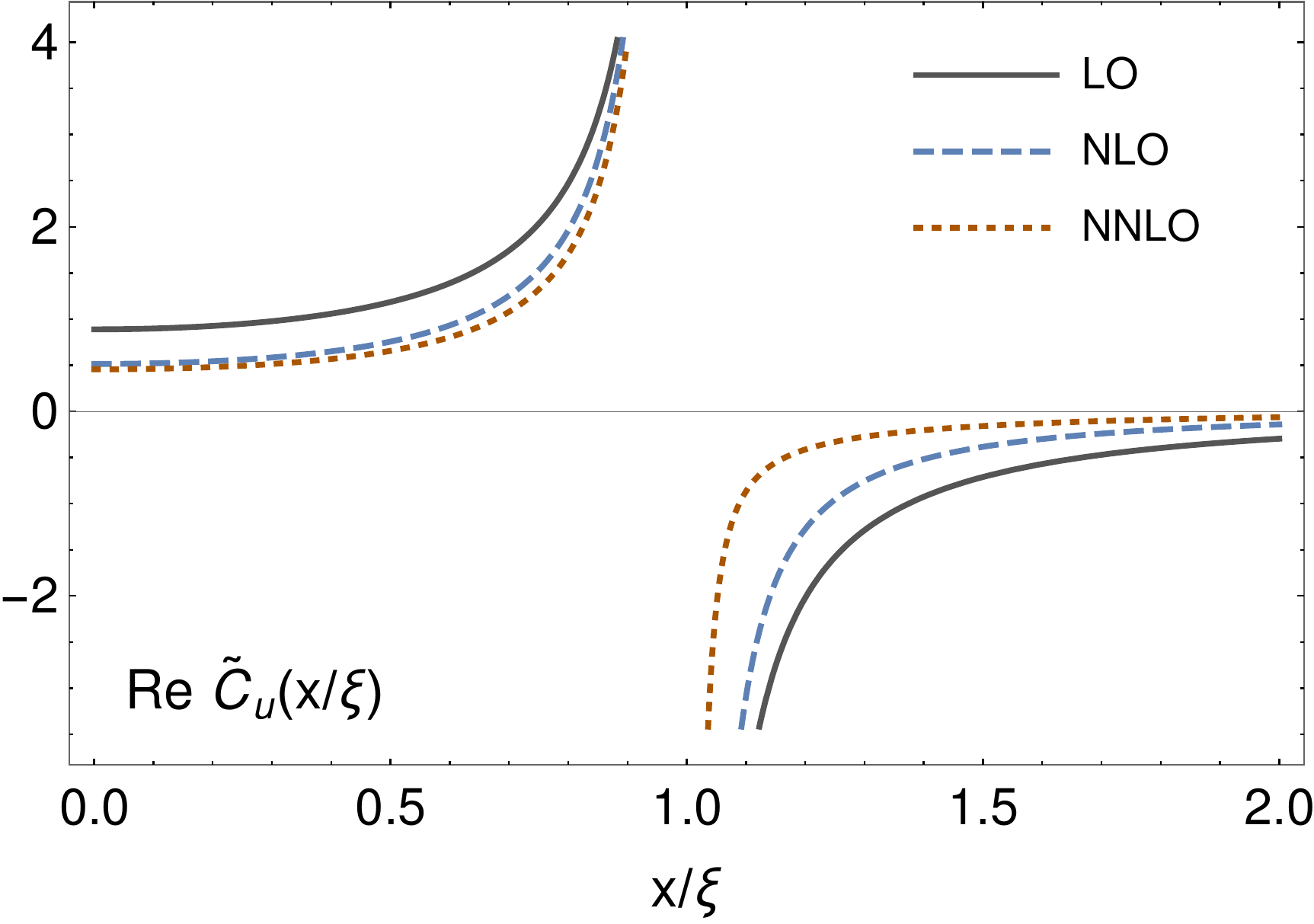} \hspace*{0.4cm}
\includegraphics[scale=0.245]{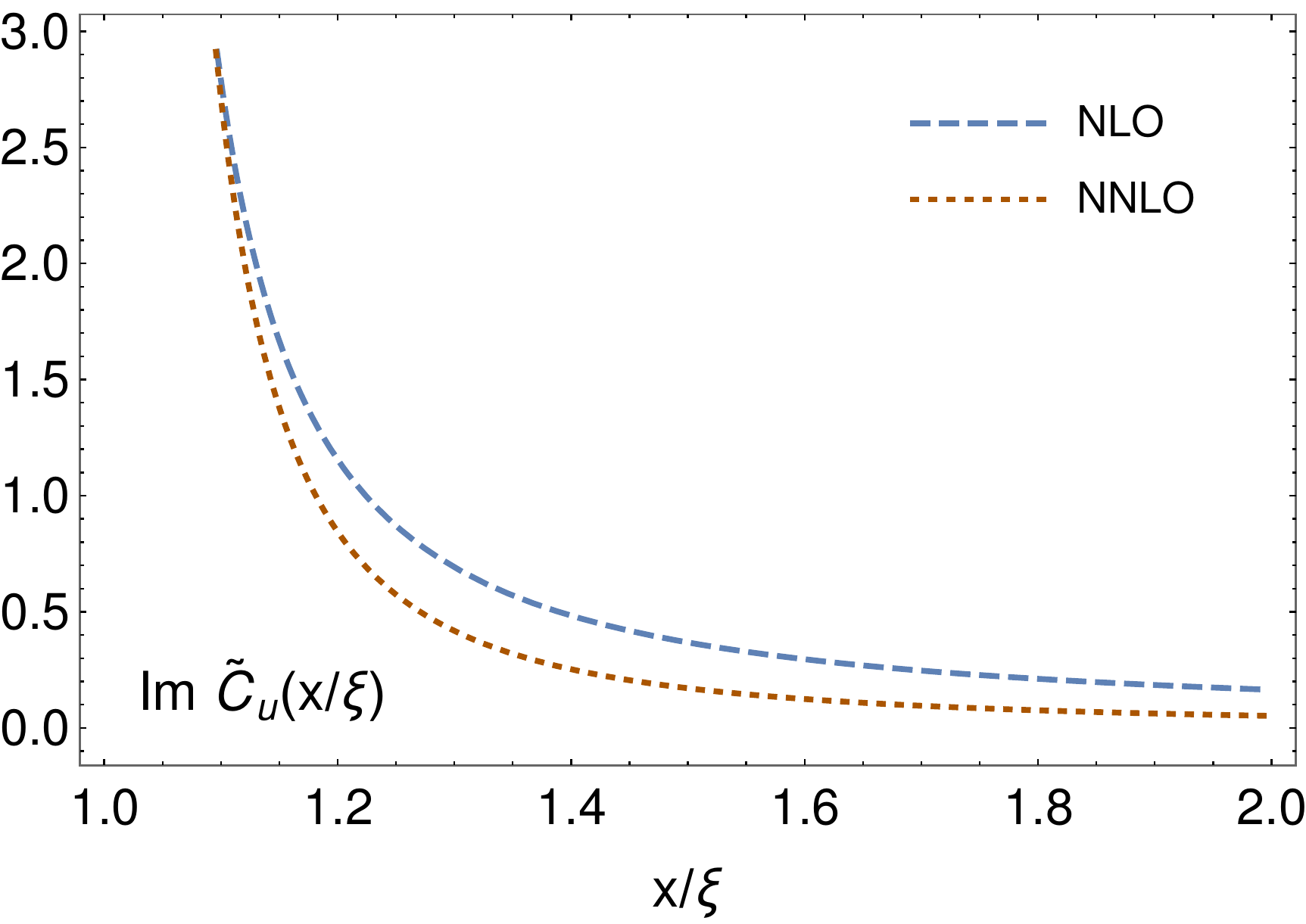} \\
\includegraphics[scale=0.245]{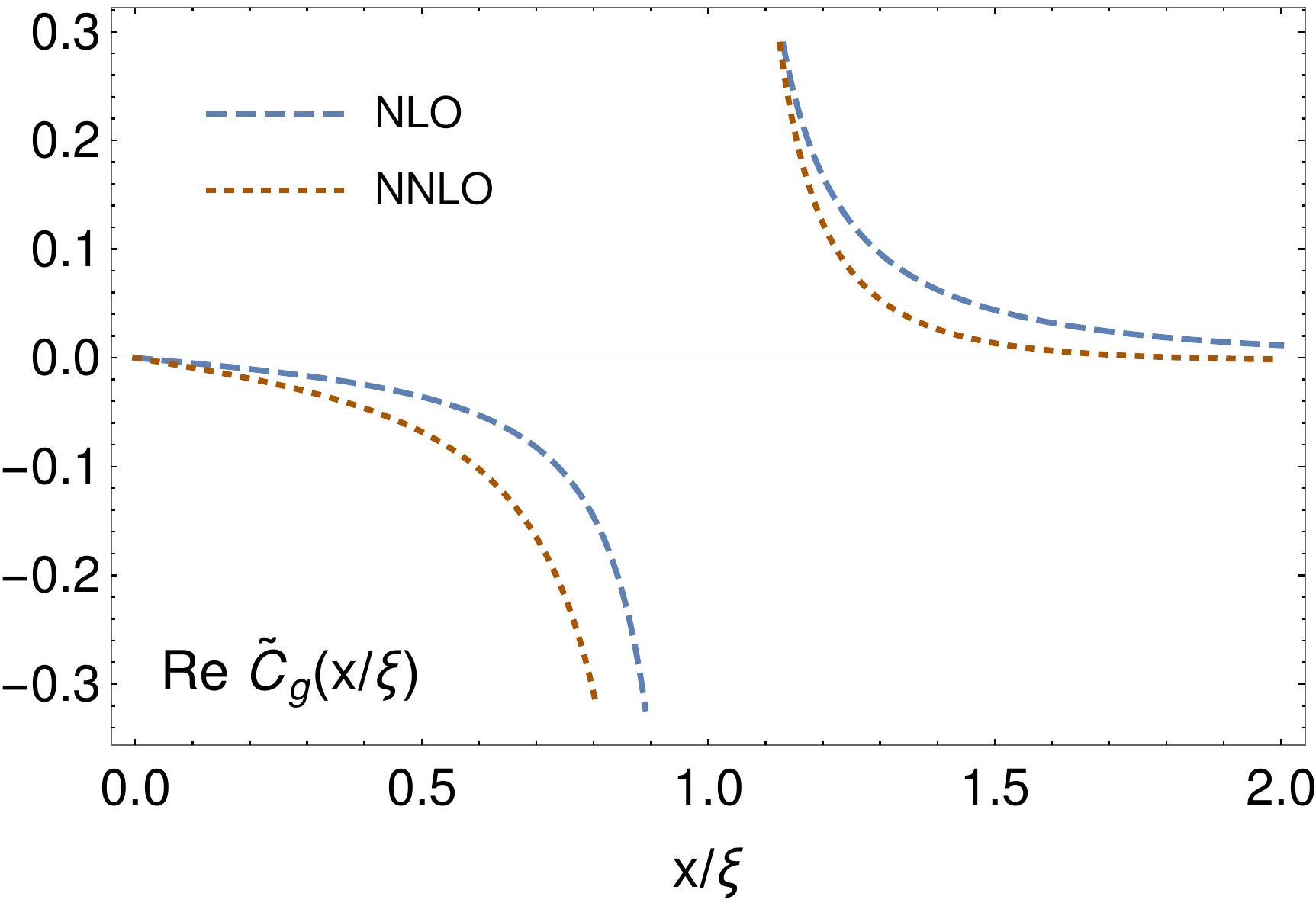} \hspace*{0.4cm}
\includegraphics[scale=0.245]{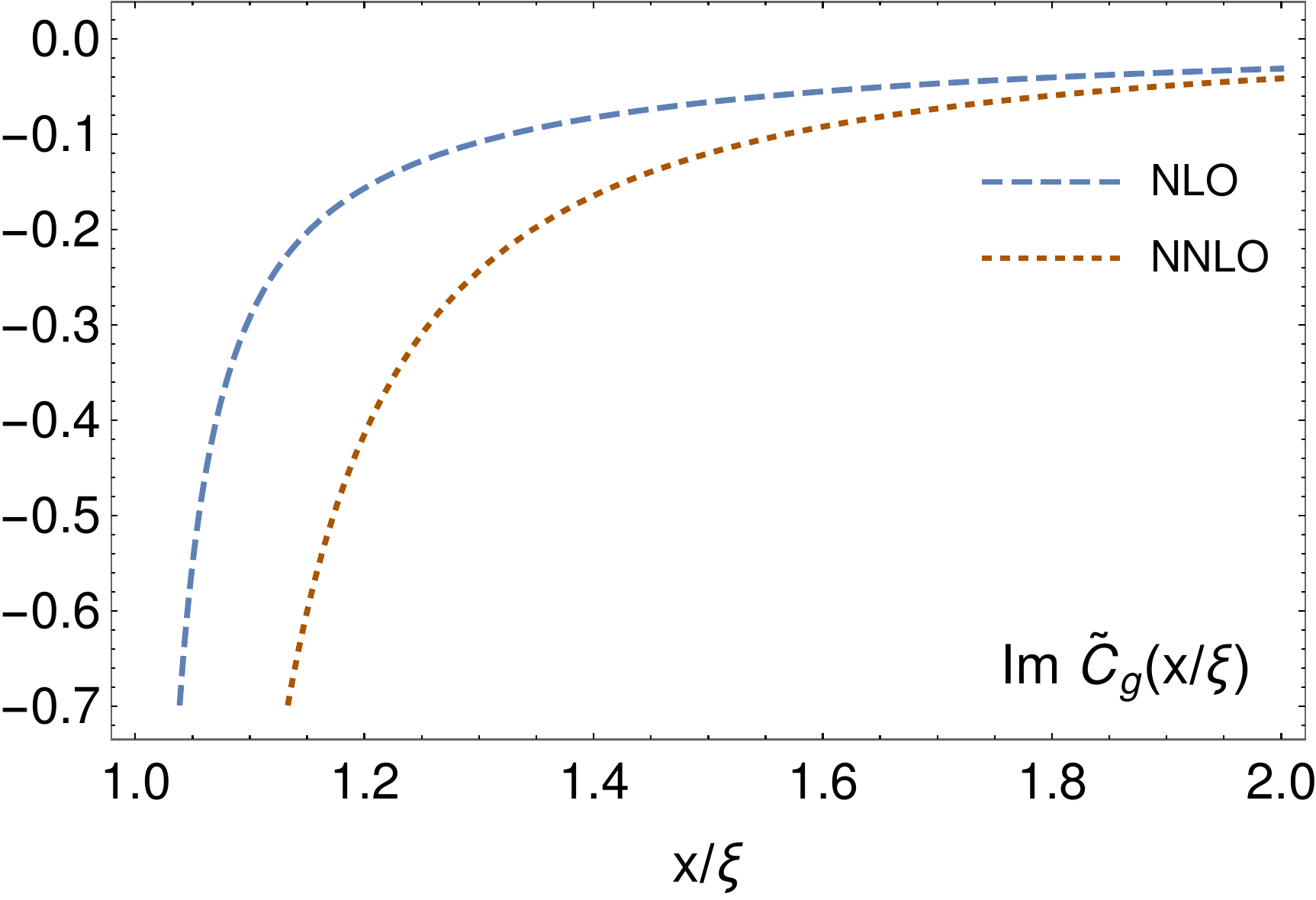} \\
\includegraphics[scale=0.245]{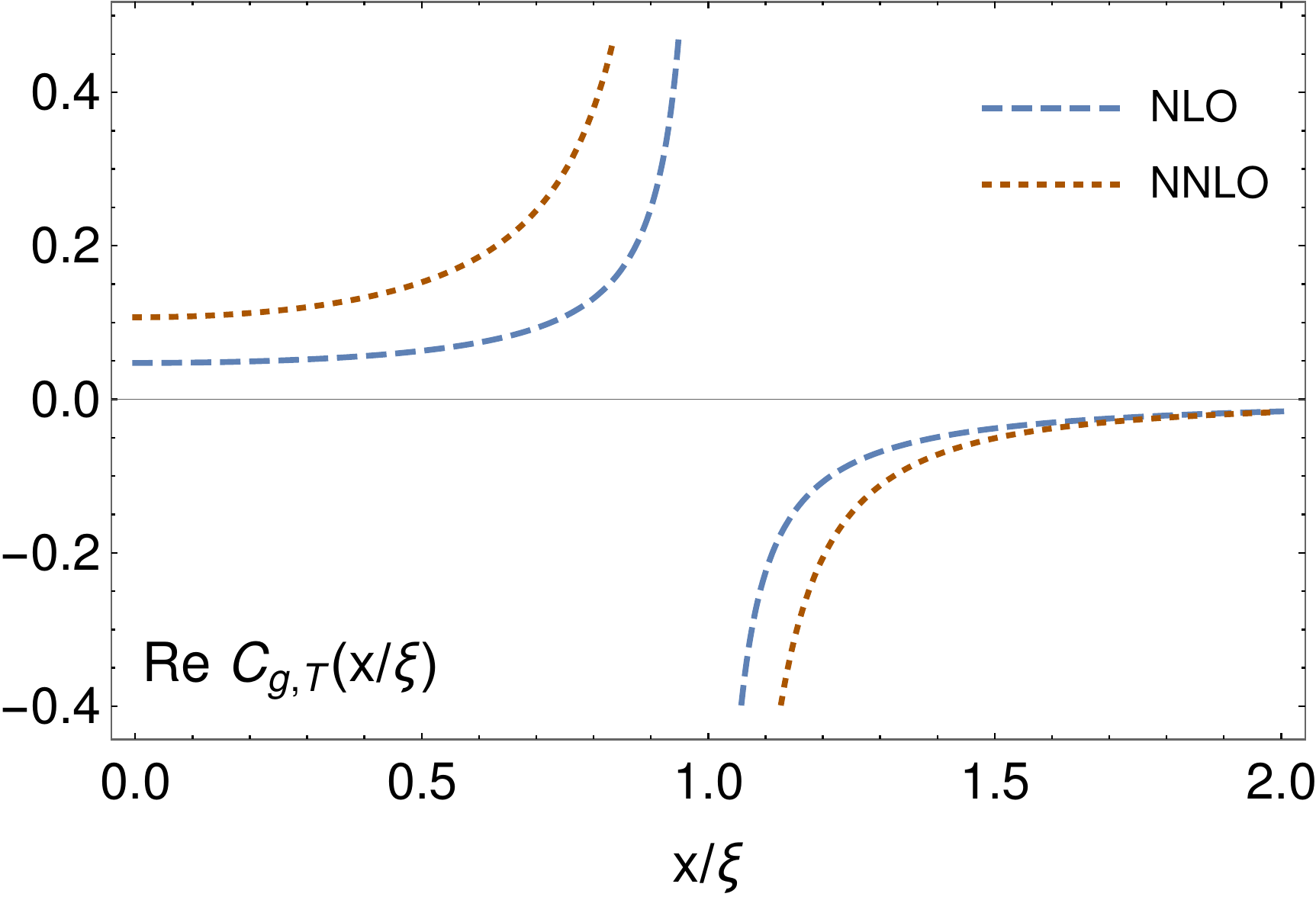} \hspace*{0.4cm}
\includegraphics[scale=0.245]{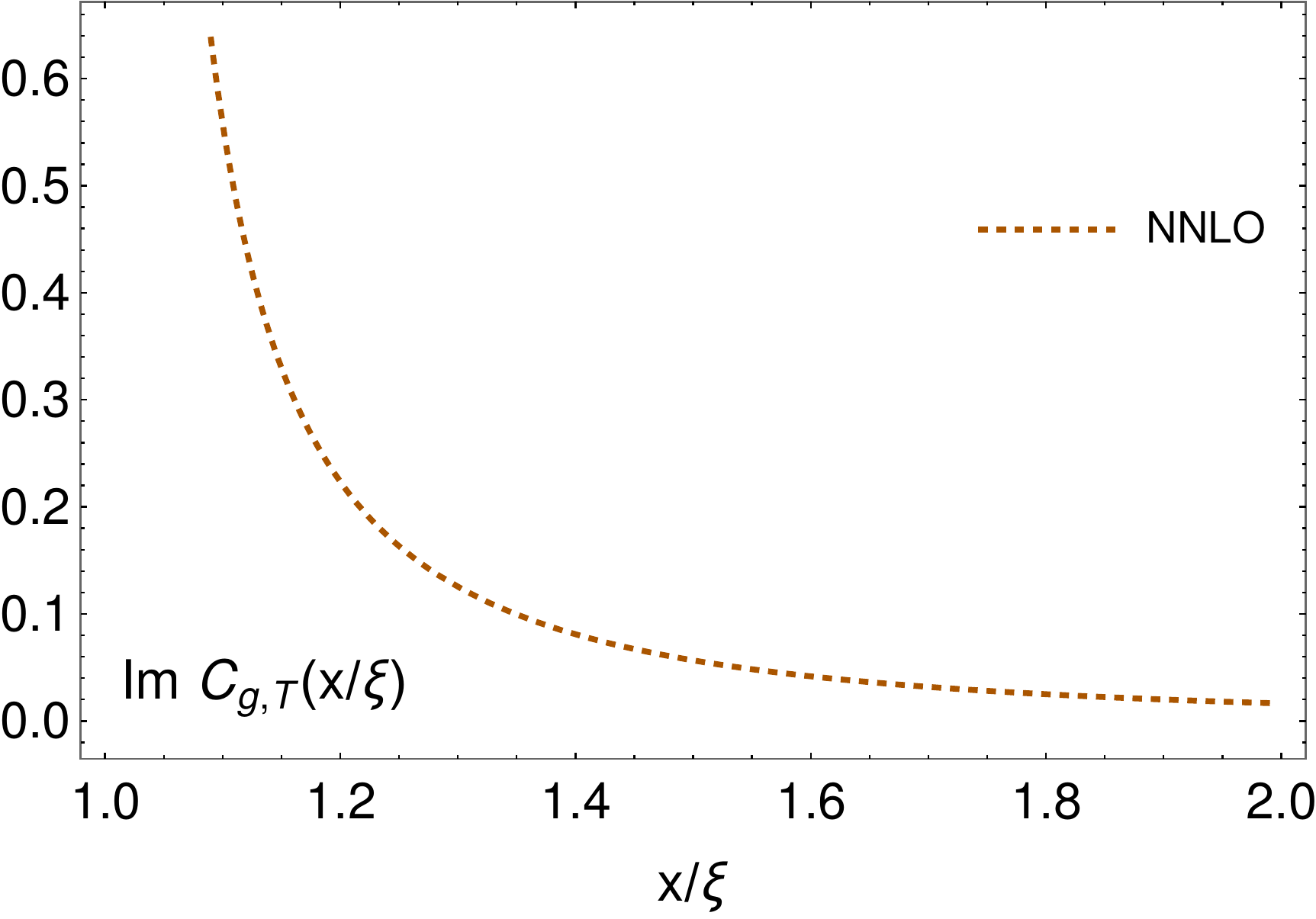}
\caption{The real (left panels) and imaginary (right panels) parts of the 
CFs $\w C_{u}, \w C_g$ and $C_{g,T}$ as functions of $x/\xi$ 
at $\mu^2 = Q^2 = 4\text{ GeV}^2$ for $n_f=3$ and $\alpha_s(4\text{ GeV}^2) = 0.29751$ for NLO, 
$\alpha_s(4\text{ GeV}^2) = 0.300224$ for NNLO.
Solid lines: LO (black), short dashes: NLO (blue), long dashes: NNLO (orange).
The delta-function contributions to the imaginary parts at $x=\xi$ 
are  not shown. 
}
\label{figCF}
\end{figure*}

This presents the main results of our paper. The vector contributions $C_q^{(2)}, C_g^{(2)}$ have been calculated in \cite{Braun:2022bpn} and we do not repeat the results here.  We organize the two-loop axial-vector and transversity contributions in the following way
\begin{align}
\w C_q^{(2)}(x/\xi, L) &=\biggl[ \frac{1}{2z \bar z}  \biggl( e_q^2 C_F^2 \w { \text C}_{\text{NS}}^{(F)}(z,L) + C_F C_A \w { \text C}_{\text{NS}}^{(A)}(z,L)  
+ C_F \beta_0 \w { \text C}_{\text{NS}}^{(\beta_0)} (z,L)  
\notag\\&\quad 
+ \Big ( \sum_{q'} e_{q'}^2 \Big ) T_F C_F \w { \text C }_{\text{PS}} (z,L) \biggr)\biggr] + [z \leftrightarrow \bar z]\label{eq: CF2quark},
\\
\w C_g^{(2)}(x/\xi, L) &= \biggl[\Big ( \sum_q e_q^2 \Big )\frac{T_F}{4z^2 \bar z^2}   \Big ( C_F \w { \text C}_{g}^{(F)} (z,L) + C_A \w { \text C}_{g}^{(A)} (z,L) \Big )\biggr] - [z\leftrightarrow \bar z]\, ,
\label{eq:CF2}
\end{align}
where ``PS" stands for `` pure-flavor singlet".
The contributions $\w { \text C}_{\text{NS}}^{(F)}, \w { \text C}_{\text{NS}}^{(A)}, \w { \text C}_{\text{NS}}^{(\beta_0)}$ have first been calculated in \cite{Braun:2021grd, Gao:2021iqq}, but the results were given only in the NDR scheme. We present our results in Larin's scheme in terms of harmonic polylogarithms (HPLs)~\cite{Remiddi:1999ew, Maitre:2005uu, Maitre:2007kp} in the following\footnote{A file with our main results is available on the preprint server {\tt{\href{http://arXiv.org} {\color{blue}http://arXiv.org}}} by downloading the source. Additionally, they can also be requested from the authors.},

\begin{align*}
\w { \text C}_{\text{NS}}^{(F)}(z,L) &= 2L^2\left(z(\Hp_2+2\Hp_{1,1})-\bar z(\Hp_1-3\Hp_0)+\frac98-\frac{\pi^2}{12}\right)
-4L\bigg(z(\Hp_3+2\Hp_{1,1,0}-2\Hp_2-3\Hp_{1,1})\notag\\
&\qquad+\bar z(2\Hp_{2,0}+3\Hp_{0,0,0})+\frac92\Hp_{0,0}-\frac{z}{4}\left(\Hp_0-\left(18+\frac43\pi^2\right)\Hp_1\right)-\frac{51}{16}+\frac{\pi^2}{24}+2\zeta_3\bigg)\notag\\
&+2z(\Hp_{3,1}+2\Hp_{3,0}-5\Hp_{2,2}-\Hp_{2,1,1}+4\Hp_{1,2,1}-6\Hp_{1,2,0}+4\Hp_{1,1,2}+6\Hp_{1,1,1,1}+4\Hp_{1,1,0,0})\notag\\
&-2\bar z(8\Hp_4-2\Hp_{1,3}-3\Hp_{2,1,0}+\Hp_{1,1,1,0})+24z\bar z\Hp_3-32z\Hp_3+8\Hp_{2,1}+24z\bar z\Hp_{1,2}\notag\\
&-22\bar z\Hp_{1,2}-18\bar z\Hp_{1,1,1}-(2+(18+\pi^2)\bar z)\Hp_2+\left(\left(\frac53\pi^2-13\right)\bar z-\left(18+\frac{4\pi^2}3\right)\right)\Hp_{1,1}\notag\\
&-\left(\left(\frac{19}{2}-\frac{7\pi^2}3-38\zeta_3\right)\bar z+\pi^2+28\zeta_3\right)\Hp_1 + 48\zeta_3z\bar z+\frac{331}{16}+\frac{7\pi^2}3-\frac{45}2\zeta_3+\frac{101}{360}\pi^4\, ,
\end{align*}

\begin{align*}
\w { \text C}_{\text{NS}}^{(A)}(z,L)  &= -4L\left(\bar z\Hp_3+z\Hp_{1,2}+\left(\frac23-\frac{5z}{3}\right)\Hp_0-\frac{\zeta_3}2+\frac18\right) + 2\bar z (3\Hp_4 -\Hp_{3,1}-\Hp_{3,0}+\Hp_{1,3})\notag\\
&+2\bar z\Hp_{2,2}-6z\Hp_{1,2,1}+4\Hp_{1,2,0}-4\Hp_{2,1,0}-2\Hp_{1,1,2}-2\Hp_{1,1,0,0}+2\Hp_{1,1,1,0}-12z\bar z\Hp_3\notag\\
&-2z(1+6z)\Hp_{2,0}-14\bar z\Hp_{1,1,0}+4\bar z\Hp_{2}-\frac{22}3\bar z\Hp_{1,0}+2\Hp_{1,1}+\left(\frac{2\bar z}{3}+\frac{\pi^2z}{3}\right)\Hp_{0,0}\notag\\
&+\left(2\pi^2z\bar z-\frac{197}9\bar z+\frac{55}{3}+4\zeta_3\right)\Hp_0-\frac{73}{24}-\frac{11\pi^2}{18}+13\zeta_3-\frac{11\pi^4}{90}\, ,
\end{align*}

\begin{align*}
\w { \text C}_{\text{NS}}^{(\beta_0)} (z,L) &= L^2\left(z \Hp_1-\frac34 \right) + L\left(2z (\Hp_{1,1} - \Hp_{2})+\left(\frac{10}3z-\bar z\right)\Hp_1-\frac{19}{4}\right) +z \Hp_3 + 2z(\Hp_{1,1,1}-\Hp_{2,1}) \notag\\
& + 2\left(2-\frac{13}{3}z\right)\Hp_2-\left(1-\frac{13}{3}z\right)\Hp_{1,1} -\frac{19}{18}\left(3-5z\right)\Hp_1-\frac{457}{48}\, ,
\end{align*}

\begin{align*}
\w { \text C}_{\text{PS}} (z,L) &= 4L^2\left(2z\Hp_2+3\bar z\Hp_1-\frac{\pi^2}{6}\right) -8L\left(z\left(2\Hp_3-2\Hp_{2,1}-3\Hp_2\right)-3\bar z\Hp_{1,1}-8\bar z\Hp_{1}+\frac{\pi^2}{4}\right)\notag\\
&+8z\left(\Hp_4-2\Hp_{3,1}+2\Hp_{2,1,1}+\left(2\bar z-\frac{11}2\right)\Hp_3+2\bar z\Hp_{1,2}+3\Hp_{2,1}+8\Hp_2\right)\notag\\
&-8\bar z\left(2\Hp_{1,2}-3\Hp_{1,1,1}-8\Hp_{1,1}-19\Hp_1\right)-\frac{16\pi^2}3+2(13-16z^2)\zeta_3-\frac{\pi^4}{9}\, ,
\end{align*}

\begin{align*}
\w { \text C}_{g}^{(F)} (z,L) &= -L^2\bar z\left(2\bar z\Hp_{1,1}-(1+z)\Hp_1\right)-4L\bigg(2\bar z^2\Hp_{1,1,1}-z^2\Hp_{2}+2\bar z(2\bar z-1)\Hp_{1,1}+\left(4-\frac{\pi^2}6\right)\bar z^2\Hp_1\notag\\
&\quad-3\bar z\Hp_1+\frac{\pi^2}{6}z\bigg) -8z^2\left(\Hp_4-\frac54\Hp_{3,0}+\Hp_{2,2}\right)-8\bar z^2\left(\Hp_{1,3}-\frac14\Hp_{1,2,1}+\frac54\Hp_{1,1,1,1}\right)\notag\\
&+z(16z^2-33z+8)\Hp_3+4z^2\Hp_{2,1}+8z\bar z(1-2z)\Hp_{1,2}+2(13z-8)\bar z\Hp_{1,1,1}+3z\left(5z-4\right)\Hp_2\notag\\
&-\left((\pi^2+30)\bar z-20\right)\bar z\Hp_{1,1}+2\left(\left(19-\frac{5\pi^2}{6}\right)z-10+\frac{2\pi^2}3+7\zeta_3\bar z\right)\bar z\Hp_1\notag\\
&-\left(\frac{\pi^2}{2}-\left(5-32z\bar z\right)\zeta_3-\frac{22\pi^4}{45}\right)z\, ,
\end{align*}

\begin{align*}
\w { \text C}_{g}^{(A)} (z,L) &= 2L^2\left(3z^2\Hp_2-\bar z^2\Hp_{1,1}-4\bar z(1-2z)\Hp_1-\frac{\pi^2}2z\right)-4L\bigg(2z^2\Hp_3-(\bar z^2+3z^2)\Hp_{2,1}\notag\\
&\quad+\bar z^2(\Hp_{1,2}+\Hp_{1,1,1})-4z^2\Hp_2+\bar z(5-9z)\Hp_{1,1}+\left(10-\frac{\pi^2}{6}\bar z-19z\right)\bar z\Hp_1+\left(\frac{2\pi^2}3+\zeta_3\right)z\bigg)\notag\\
&+4z^2\left(\Hp_4-\Hp_{3,0}\right)-2(\bar z^2+2z^2)\left(2\Hp_{3,1}-3\Hp_{2,1,1}\right)+4\bar z^2\left(\Hp_{2,2}+\Hp_{1,3}-\Hp_{1,2,1}-\frac12\Hp_{1,1,1,1}\right)\notag\\
&-2z(2+5z+4z^2)\Hp_3+2(3-7z+12z^2)\Hp_{2,1}-4z^2(1-2z)\Hp_{2,0}-2\bar z(9-16z)\Hp_{1,1,1}\notag\\
&-z(13-46z)\Hp_2-\bar z\left(46-\frac{4\pi^2}{3}\bar z-80z\right)\Hp_{1,1}-\bar z\left(84-\frac{4\pi^2}3\bar z^2-4\zeta_3-4z(40-\zeta_3)\right)\Hp_1\notag\\
&-\left(\frac{11\pi^2}{2}-14\zeta_3+\frac{13\pi^4}{45}\right)z\, ,
\end{align*} 
with the argument $z$ for all HPL functions $\Hp_{\vec m}$ omitted for brevity. 
In Figure \ref{figCF} we have plotted the real and imaginary parts of $\w C_q$ and $\w C_g$ as a function of $x/\xi$. 
The higher order corrections look almost identical to the vector case \cite{Braun:2022bpn}. We therefore expect the corrections to the Compton form factors $\w {\f H}, \w {\f E}$ to be of the same size as the corrections to $\f H, \f E$, which were studied in \cite{Braun:2022bpn} using the Goloskokov-Kroll model \cite{Goloskokov:2006hr} for the unpolarized GPDs $H_q$ and $H_g$. A more detailed study is left for future work.

We remark that, just as at NLO, the axial-vector CF and  vector CF have identical leading terms as $z \rightarrow 0$, that is
\begin{align}
\w C_q^{(2)} &\simeq C_q^{(2)} \simeq \frac{e_q^2 C_F^2}{4z} \log^4z, \nonumber
\\
\w { \text C }_{\text{PS}} &\simeq { \text C }_{\text{PS}} \simeq - \frac{2}{3} \log^3z,
\\
\w C_g^{(2)} &\simeq C_g^{(2)} \simeq T_F (C_A + 5C_F) \frac{\sum_q e_q^2}{12z} \log^3z. \nonumber
\end{align}

\subsection*{Two-loop gluon transversity CF}
The two-loop gluon transversity CF reads
\begin{align}
C_{g,T}^{(2)}(x/\xi, L) &= \Big (  \sum_q e_q^2 \Big )  \frac{T_F}{z \bar z^2}  \Big [ - C_F \Big (\log z + 2 \bar z\Big ) + C_A \Big (\log^2 z - (5+2L) \log z + \bar z \Big )\Big ]  \nonumber
\\
&\quad + [z \leftrightarrow \bar z].
\end{align}
Apparently, this function has transcendental weight reduced by two compared to the vector and axial-vector sectors. 
From the perspective of loop calculations, such a reduction in complexity, both in transcendentality and the form of the expression, is highly nontrivial considering the fact that it is necessary to compute the same set of Feynman diagrams (with  Lorentz projector in~\eqref{eq: Eperp def}) for $C_{g,T}^{(2)}$ as for the vector and axial-vector case. The diagrams individually may have nonvanishing contributions to the transversity CF up to transcendental weight four. In other words, the reduction in the weight only occurs when all two-loop diagrams are summed up.
At two-loops, this can be seen without any explicit calculation, as a consequence of properties of the two-loop master integrals and the factorization theorem, which guarantees the cancellation of IR singularities in the CF. See Appendix \ref{sec: B} for a detailled argument.

For a numerical estimation, we have plotted in Figure \ref{figCF} the real and imaginary parts of $C_{g,T}$ on the real axis. The results are as expected for a first-order correction.

\section{Conclusion}
We have presented the two-loop coefficient functions responsible for the flavor singlet axial-vector and gluon transversity contributions in DVCS, which in combination with the previous publications, completes the full NNLO CFs responsible for the leading-power contributions in DVCS. 

The calculations are analogous to the vector case, which was considered in \cite{Braun:2022bpn}. The interesting aspects are in the use of dimensional regularization. In terms of the decomposition of the transverse photon indices, the most straightforward continuation given in eq.~\eqref{eq: perp decomp} was used. The unavoidable ambiguity concerns the treatment of $\gamma_5$. For the calculations in this paper we have employed the Larin's scheme and explained how the result may be converted to the NDR scheme in Section \ref{sec: Larin to MSbar}. We leave the task of making such a conversion for future work.

Our plots of the coefficient functions at varying orders of $\alpha_s$ were shown in Figure \ref{figCF}. With regards to the NNLO corrections, no new features compared to the vector case can be observed. We therefore expect the size estimations for the NNLO corrections to the corresponding Compton form factors to be analogous to \cite{Braun:2022bpn} for particular DVCS channels. A dedicated phenomenological study, including the theory uncertainties in the form of the scale dependence of the Compton form factors, is required to fully assess the NNLO impact in the current (JLAB-12) and future (EIC) experiments. We leave this as a future project.

We observe that, up to two loops, $C_{g,T}$ has transcendental weight reduced by two compared to the vector and axial-vector case. In Appendix \ref{sec: B} we provide an explanation for the weight reduction up to two-loops. It would be interesting to see if such reductions hold to higher orders.

\begin{acknowledgments}
This work was supported in part by the Research Unit FOR2926 and 
the Collaborative Research Center TRR110/2 funded by 
the Deutsche Forschungsgemeinschaft (DFG, German Research Foundation) under grants 409651613 and 
196253076, respectively.     
\end{acknowledgments}
%

\appendix

\section{Some details of the calculation}
\label{sec: A}

To calculate the CFs using the matching procedure we must make the various choices of external states that project on the desired structure. For the entirety of this section, the leading twist approximation $m^2 = t = 0$ is implied unless otherwise stated. 
To be precise, we distinguish now bare and renormalized fields. In particular, we clarify that for the definition of the hadronic tensor in eq. \eqref{eq: hadronic tensor} the electromagnetic currents are bare, i.e. they are composite operators of bare fields without any renormalization constant. By current conservation this is the same as the renormalized electromagnetic current
\begin{align}
j^{\mu} = \sum_q e_q^2 \bar \psi_q^{\rm bare} \gamma^{\mu} \psi_q^{\rm bare} = \sum_q e_q^2  Z_{J,q}\bar \psi_q^{\rm R} \gamma^{\mu} \psi_q^{\rm R},
\end{align}
where $Z_{J,q} = Z_q$, with $Z_q$ being the quark wave-function renormalization constant $\psi_q^{\rm bare} = \sqrt{Z_q} \psi_q^{\rm R}$, is a standard result.
On the other hand, the GPDs are strictly given in terms of renormalized composite bilocal operators of renormalized fields.

\subsection*{External parton states}
For the quark vector case, we choose
\begin{align}
\bra{\text{out}} (...) \ket{\text{in}} &\rightarrow \lim_{p \rightarrow p^+ \bar n, p'^+ \rightarrow p'^+ \bar n} \int d^dx \, e^{-ipx}\int d^dy \, e^{ip'y} \frac{1}{N_c}  \sum_j \frac{P^+}{2} (-i)^2 (\s p \gamma^-\s p')_{\beta \alpha} \nonumber
\\
&\quad \times  \langle \Omega | T \{ q_{\alpha, j}^{\rm bare}(y) (...) \bar q_{\beta, j}^{\rm bare}(x) \} | \Omega \rangle_c,
\label{eq: vector quark ext states}
\end{align}
where $\alpha, \beta$ are Dirac indices, $j$ is a color index and $\langle \Omega |T \{ ... \} | \Omega \rangle _c$ denotes the sum of connected Feynman diagrams. Note that the leading twist limit should only be taken after the amputation by the factors of $-i \s p$ and $-i\s p'$. 

With the external states in eq. \eqref{eq: vector quark ext states} we find
\begin{align}
\widehat{F}_{q/q'}(x,\xi) &= \delta_{qq'} \delta(1-x) + O(\alpha_s).
\label{eq: Fqq0}
\end{align}

For the gluon vector case we choose
\begin{align}
\bra{\text{out}} (...) \ket{\text{in}} &\rightarrow \lim_{p \rightarrow p^+ \bar n, p' \rightarrow p'^+ \bar n} \int d^dx \, e^{-ipx}\int d^dy \, e^{ip'y} \frac{ g_{\perp, \mu \nu} }{2(d-2)}\frac{\delta^{ab}}{8} (ip^2)(ip'^2) \nonumber
\\
&\quad \times \langle \Omega | T \{ A^{\nu, b, \rm bare} (y) (...) A^{\mu, a, \rm bare}(x) \} |\Omega \rangle_c.
\label{eq: vector gluon ext states}
\end{align}
This gives
\begin{align}
\widehat{F}_{g/g}(x,\xi) &= (\xi^2 - 1) \frac{1}{2} \Big ( \delta(1- x) + \delta(1+x) \Big ) + O(\alpha_s).
\label{eq: Fgg0}
\end{align}
For the axial-vector and transversity case the situation is different, since we have two free Lorentz indices. When choosing external states we will introduce two additional Lorentz indices. Then the resulting factorization formulas in eq. \eqref{eq: factorization partonic} can be projected with the corresponding rank $4$ tensor, $\chi_{\perp}$ for axial-vector, $\tau_{\perp}$ for transversity, giving a factorization in terms of scalar quantities.

For axial-vector we take
\begin{align}
\bra{\text{out}} (...)^{\mu \nu} \ket{\text{in}} &\rightarrow \lim_{p \rightarrow p^+ \bar n, p' \rightarrow p'^+ \bar n} \int d^dx \, e^{-ipx}\int d^dy \, e^{ip'y} \frac{1}{N_c}  \sum_j \frac{P^+}{2} (-i)^2 \nonumber 
\\
&\quad \times \frac{ i\chi_{\perp, \mu \nu \mu' \nu'} }{(d-3)(d-2)} (\s p \Gamma^{-\mu' \nu'}\s p')_{\beta \alpha}  \langle \Omega | T \{ q_{\alpha, j}^{\rm bare}(y) (...)^{\mu \nu} \bar q_{\beta, j}^{\rm bare}(x) \} | \Omega \rangle_c,
\label{eq: axial quark ext states}
\\
\bra{\text{out}} (...)^{\mu \nu} \ket{\text{in}} &\rightarrow \lim_{p \rightarrow p^+ \bar n , p' \rightarrow p'^+ \bar n} \int d^dx \, e^{-ipx}\int d^dy \, e^{ip'y} \frac{ i\chi_{\perp, \mu \nu \mu' \nu'} }{2(d-3)(d-2)} \nonumber
\\
&\quad \times \frac{\delta^{ab}}{8} (ip^2)(ip'^2) \langle \Omega | T \{ A^{\nu', b, \rm bare} (y) (...)^{\mu \nu} A^{\mu', a, \rm bare}(x) \} | \Omega \rangle_c
\label{eq: axial gluon ext states}
\end{align}
for the quark and gluon case respectively. This gives
\begin{align}
\widehat{\w F}_{q/q'}(x,\xi) &= \delta_{qq'} \delta(1-x) + O(\alpha_s),
\label{eq: hatFqq0}
\\
\widehat{\w F}_{g/g}(x,\xi) &= (\xi^2 - 1) \frac{1}{2} \Big ( \delta(1- x) - \delta(1+x) \Big ) + O(\alpha_s),
\label{eq: hatFgg0}
\end{align}
where now there are no free indices compared to eq. \eqref{eq: axial GPDs partonic}, because they were projected out in the prescriptions in eqs. \eqref{eq: axial quark ext states} and \eqref{eq: axial gluon ext states}.

For the transversity case we choose
\begin{align}
\bra{\text{out}} (...)^{\mu \nu} \ket{\text{in}} &\rightarrow \lim_{p \rightarrow p^+ \bar n, p' \rightarrow p'^+ \bar n} \int d^dx \, e^{-ipx}\int d^dy \, e^{ip'y} \frac{ \tau_{\perp, \mu \nu \mu' \nu'} }{d(d-3)} \nonumber
\\
&\quad \times \frac{\delta^{ab}}{8} (ip^2)(ip'^2) \langle \Omega | T \{ A^{\nu', b, \rm bare} (y) (...)^{\mu \nu} A^{\mu', a, \rm bare}(x) \} |\Omega \rangle_c,
\label{eq: transversity ext states}
\end{align}
which leads to
\begin{align}
\widehat F_{g/g,T} (x,\xi) &= (\xi^2 - 1) \frac{1}{2} \Big ( \delta(1-x) +  \delta(1+x)\Big ) + O(\alpha_s).
\label{eq: FggT0}
\end{align}

\subsection*{UV divergences}
The partonic tensor $\widehat T_f$ (dropping the Lorentz indices for simplicity), amputated with the prescriptions eqs. \eqref{eq: vector quark ext states}, \eqref{eq: vector gluon ext states}, \eqref{eq: axial quark ext states}, \eqref{eq: axial gluon ext states}, \eqref{eq: transversity ext states} is formally UV divergent, since we defined the Greens function in terms of bare fields. To make it UV finite we should multiply by inverse wave-function renormalization $Z$-factors to turn the bare into renormalized fields.

We are however only interested in the CFs, which do not care about the external states. After all, multiplying by the wave-function renormalization factors corresponds to multiplying both sides of eq. \eqref{eq: factorization partonic} by the same constant. In fact, since we do not need to distinguish between $\epsilon_{\rm UV}$ and $\epsilon_{\rm IR}$, it is more convenient to not multiply by $Z$-factors. Any external leg correction is a vanishing scaleless integral and therefore the residues in the poles of the external propagators at $\s p,\s p' = 0$ are unity.

This leaves us only with the renormalization of the coupling constant
\begin{align}
\alpha_s^{\rm bare} = \alpha_s \mu^{2\epsilon} \Big ( 1 - \frac{\alpha_s \beta_0}{4\pi \epsilon} + O(\alpha_s^2) \Big ).
\end{align}
Expanding the partonic tensor in terms of both the unrenormalized and renormalized coupling gives
\begin{align}
\widehat T_f &= 1 + \frac{\alpha_{s}^{\rm bare}}{4\pi} \widehat T_f^{(1), \rm bare} + \Big (\frac{\alpha_{s}^{\rm bare}}{4\pi} \Big )^2 \widehat T_f^{(2), \rm bare} + O((\alpha_s^{\rm bare})^3) \nonumber
\\
&= 1 + \frac{\alpha_{s}}{4\pi} \widehat T_f^{(1)} + \Big ( \frac{\alpha_{s}}{4\pi} \Big )^2 \widehat T_f^{(2)} + O(\alpha_s^3).
\end{align}
Hence
\begin{align}
\widehat T_f^{(1)} &= \mu^{2\epsilon} \widehat T_f^{(1),\rm bare}, \nonumber
\\
\widehat T_f^{(2)} &= \mu^{4\epsilon} \widehat T_f^{(2), \rm bare} - \frac{1}{\epsilon} \beta_0 \widehat T_f^{(1)}.
\end{align}
In particular, there is a finite subdivergence subtraction term, which is given by $- \beta_0 \widehat T_f^{(1,1)}$. 

We remark that although such a subtraction is present also in the gluon case, the two-loop gluon CF turns out to not depend on $\beta_0$. This is because in the matching there is an additional contribution proportional to $\beta_0$, due to the gluon GPD being defined in terms of renormalized gluon fields, which exactly cancels this term. We will come back to this below, see eq. \eqref{eq: beta0 cancellation}.

\subsection*{Partonic GPDs}

Without renormalization, the GPD with partonic on-shell external states would be given exactly by eqs. \eqref{eq: Fqq0}, \eqref{eq: Fgg0}, \eqref{eq: hatFqq0}, \eqref{eq: hatFgg0}, \eqref{eq: FggT0} without the $O(\alpha_s)$ terms, since all loop integrals are scaleless. Furthermore, the off-diagonal partonic GPDs would be zero. 
Formally, this means that the UV and IR singularities ``cancel out''. But strictly speaking, the factorization formula in eq. \eqref{eq: factorization} is formulated in terms of the renormalized GPDs.

After renormalizing the GPDs, only the IR singularities remain, giving purely IR divergent higher order terms to the partonic GPDs. The renormalization $Z$-factor of the GPDs can be reconstructed from the evolution kernels, which are known to two-loop accuracy \cite{Belitsky:1999hf, Belitsky:2000jk}. A complete list of the one-loop evolution kernels in position space can be found in \cite{Belitsky:2005qn}, section 4.3.2. 

At one-loop we have
\begin{align}
Z_{ff'}^{(1)} = \frac{1}{2\epsilon} \f K_{ff'}^{(1)},
\end{align}
where $\f K_{ff'}$ are the evolution kernels for the mixing of a parton of species $f$ to a parton of species $f'$. Note that $Z_{ff'}$ defined here includes a factor of $Z_A^{-1}$ for $f' = g$ and $Z_q^{-1}$ for $f' = q$, which comes from the GPD being defined in terms of renormalized fields, such that
\begin{align}
\f O_f^{\rm R} = \sum_{f'} Z_{ff'} \ast \f O_{f'}^{\rm bare}.
\end{align}
Upon identifying $\epsilon = \epsilon_{\rm IR} = \epsilon_{\rm UV}$ the partonic matrix element of $\f O_{f'}^{\rm bare}$ is given by the tree-level expression, so $\f O_f^{\rm R}$ is entirely determined by $Z_{ff'}$.

We list the complete set of one-loop partonic GPDs in terms of position space kernels. For notational simplicity we let
\begin{align}
\bar \alpha = 1-\alpha, \qquad  \bar \beta = 1-\beta
\end{align}
and
\begin{align}
\omega(\alpha, \beta)=  \frac{1+\xi}{2} (1-2\alpha) + \frac{1-\xi}{2} (1-2\beta).
\end{align}
$\bullet$  Vector
\begin{align}
\widehat F_{q/q}^{(1)}(x,\xi) &=  \frac{C_F}{\epsilon} \Big \{ \delta(x-1) - 2 \int_0^1 d\alpha \int_0^{\bar \alpha} d\beta \,  \delta (x - \omega( \alpha, \beta)) \nonumber
\\
&\quad + 2\int_0^1 d\alpha \, \frac{\bar \alpha}{\alpha} \Big [  2 \delta(x - 1) - \delta ( x - \omega(\alpha, 0))
 - \delta ( x - \omega(0, \alpha) ) \Big ] \Big \}, \nonumber
\\
\widehat F_{q/g}^{(1)}(x,\xi) &= - \frac{T_F}{\epsilon} (\xi^2- 1) \int_0^1 d\alpha \int_0^{\bar \alpha} d\beta \, (\bar \alpha \bar \beta + 3 \alpha \beta) \Big [ \delta'( x - \omega(\alpha, \beta))  + \delta'( x + \omega( \alpha, \beta) ) \Big ],  \nonumber
\\
\widehat{ F}_{g/q}^{(1)}(x,\xi) &= - \frac{4C_F}{\epsilon} \int_0^1 d\alpha \int_0^{\bar \alpha} d\beta \, \int_{-1}^1d\lambda \, \Big [ \omega(\alpha, \beta) \delta(x- \lambda \, \omega(\alpha, \beta)) + \delta(x-\lambda) \Big ],
\label{eq: vector partonic GPD}
\\
\widehat F_{g/g}^{(1)}(x,\xi) &=  \frac{C_A}{\epsilon} (\xi^2 - 1) \frac{1}{2} \Big \{ \Big ( 6 - \frac{\beta_0}{C_A} \Big ) \delta(x-1)  - 8 \int_0^1 d\alpha \int_0^{\bar \alpha} d\beta \, (\bar \alpha \bar \beta + 2\alpha \beta)  \delta (x - \omega( \alpha, \beta)) \nonumber
\\
&\quad + 2\int_0^1 d\alpha \, \frac{\bar \alpha^2}{\alpha} \Big [  2 \delta(x - 1) - \delta(x-\omega(\alpha, 0)) - \delta(x - \omega(0, \alpha)) \Big ]
+ (x \rightarrow- x)  \Big \}. \nonumber
\end{align} 
$\bullet$  Axial-vector
\begin{align}
\widehat {\w F}_{q/q}^{(1)}(x,\xi) &=  \widehat F_{q/q}^{(1)}(x,\xi), \nonumber
\\
\widehat {\w F}_{q/g}^{(1)}(x,\xi) &= \frac{T_F}{\epsilon} (\xi^2- 1) \int_0^1 d\alpha \int_0^{\bar \alpha} d\beta \, (1-\alpha - \beta) \Big [ \delta'( x - \omega( \alpha, \beta)) - \delta'( x + \omega( \alpha, \beta) ) \Big ],  
\label{eq: axial partonic GPD}
\\
\widehat{ {\w F}}_{g/q}^{(1)}(x,\xi) &= - \frac{4 C_F}{\epsilon} \int_0^1 d\alpha \int_0^{\bar \alpha} d\beta \, \int_1^{\infty} d\lambda \, \Big [\omega(\alpha, \beta) \delta( x- \lambda \omega(\alpha, \beta)) - \delta(x - \lambda) - (x \rightarrow -x) \Big ],\nonumber
\\
\widehat {\w F}_{g/g}^{(1)}(x,\xi) &=  \frac{C_A}{\epsilon} (\xi^2 - 1) \frac{1}{2} \Big \{ \Big ( 6 - \frac{\beta_0}{C_A} \Big ) \delta(x-1)  - 8 \int_0^1 d\alpha \int_0^{\bar \alpha} d\beta \, (1-\alpha - \beta)  \delta (x - \omega( \alpha, \beta)) \nonumber
\\
&\quad + 2\int_0^1 d\alpha \, \frac{\bar \alpha^2}{\alpha} \Big [  2 \delta(x - 1) - \delta(x-\omega(\alpha, 0)) - \delta(x - \omega(0, \alpha)) \Big ]
- (x \rightarrow- x)  \Big \}. \nonumber
\end{align} 
$\bullet$  Gluon transversity
\begin{align}
\widehat F_{g/g, T}^{(1)}(x,\xi) &=  \frac{C_A}{\epsilon} (\xi^2 - 1) \frac{1}{2} \Big \{ \Big ( 6 - \frac{\beta_0}{C_A} \Big ) \delta(x-1)  
\label{eq: transversity partonic GPD}
\\
&\quad + 2\int_0^1 d\alpha \, \frac{\bar \alpha^2}{\alpha} \Big [  2 \delta(x - 1) - \delta(x-\omega(\alpha, 0)) - \delta(x - \omega(0, \alpha)) \Big ]
+ (x \rightarrow- x)  \Big \}. \nonumber
\end{align}

\subsection*{Double-counting subtractions}

To illustrate the matching procedure, consider as an example the quark vector case. The matching for axial-vector and gluon transversity sector are completely analogous.

The tree-level factorization formula gives
\begin{align}
\widehat{\f V}_q^{(0)}(\xi) = e_q^2 \Big ( \frac{1}{\xi - 1} - \frac{1}{\xi + 1} \Big ) = \int \frac{dx}{\xi} C_q^{(0)}(x/\xi) \widehat F_{q/q}^{(0)}(x,\xi) = \frac{1}{\xi} C_q^{(0)}(1/\xi)
\end{align}
and therefore
\begin{align}
C_q^{(0)}(x) = \frac{1}{x} \widehat{\f V}_q^{(0)}(1/x) = e_q^2 \Big ( \frac{1}{1-x} - \frac{1}{1+x} \Big )
\end{align}
Going further to one-loop we obtain
\begin{align}
 C_q^{(1)}(x, L, \epsilon) &= \frac{1}{x} \sum_{n = 0}^{\infty} \epsilon^n \widehat{\f V}_q^{(1,n)}(1/x, L)  \nonumber
\\
&\quad +  \frac{1}{\epsilon} \Big ( \frac{1}{x} \widehat{\f V}_q^{(1,-1)}(1/x, L) -  \int_{-1}^1 dy \, C_q^{(0)}(xy) \widehat F_{q/q}^{(1,-1)}(y,1/x) \Big ).
\label{eq: oneloop matching}
\end{align}
The term proportional to $\frac{1}{\epsilon}$ must vanish, making $C_q^{(1)}$ finite as $\epsilon \rightarrow 0$. Note however that $C_q^{(1)}$ remains with an analytic dependence on $\epsilon$.

Going to two-loops we obtain
\begin{align}
C_q^{(2)}(x, L, \epsilon) &= \frac{1}{x} \widehat{\f V}_q^{(2,0)}(1/x, L) - \int_{-1}^1 dy \, C_q^{(1,1)}(xy, L) \widehat F_{q/q}^{(1,-1)}(y,1/x) \nonumber
\\
&\quad - x \int_{-1}^1 dy \, C_g^{(1,1)}(xy, L) \widehat F_{g/q}^{(1,-1)}(y,1/x) + O(\epsilon),
\label{eq: quark two-loop matching}
\end{align}
where we used that all IR poles must cancel. Note that there are finite double-counting subtractions proportional to $\epsilon^1$ coefficients of the one-loop CFs, $C_q^{(1,1)}, C_g^{(1,1)}$. It is therefore important to keep in mind that CF is not simply the finite piece of the parton-level amplitude. This is true at one-loop, but not beyond. Furthermore, one should not forget that in $\widehat{\f V}_q^{(2,0)}$ there is a UV subtraction term $- \beta_0 \widehat{\f V}_q^{(1,1)}$. 

For the gluon case we have
\begin{align}
C_g^{(1)}(x,L,\epsilon) &= \sum_{n = 0}^{\infty} \epsilon^n \frac{\widehat{\f V}_g^{(1,n)}(1/x,L)}{1-x^2},
\label{eq: gluon one-loop matching}
\\
C_g^{(2)}(x,L,\epsilon) &= \frac{\widehat{\f V}_g^{(2,0)}(1/x,L)}{1-x^2} - \frac{x^2}{1-x^2} \int_{-1}^1 dy\, C_g^{(1,1)}(xy,L) \widehat F_{g/g}^{(1,-1)}(y,1/x) \nonumber
\\
&\quad - \frac{x}{1-x^2} \int_{-1}^1 dy \, C_q^{(1,1)}(xy,L) \widehat F_{q/g}^{(1,-1)}(y,1/x)  + O(\epsilon).
\label{eq: gluon two-loop matching}
\end{align}
There is also a UV subtraction term $- \beta_0 \widehat{\f V}_g^{(1,1)}$ contributing to $\widehat{\f V}_g^{(2,0)}$. Consider however the contribution proportional to $\beta_0$ in the second term of the right-hand-side of eq. \eqref{eq: gluon two-loop matching}. It reads
\begin{align}
&- \frac{x^2}{1-x^2} \int_{-1}^1 dy\, C_g^{(1,1)}(xy,L) (-\beta_0) (1/x^2 - 1) \frac{1}{2} \Big ( \delta(1-y) + \delta(1+y) \Big ) \nonumber
\\
&= \beta_0 C_g^{(1,1)}(xy,L) = \beta_0 \frac{\widehat{\f V}_g^{(1,1)}(1/x,L)}{1-x^2}.
\label{eq: beta0 cancellation}
\end{align}
We conclude that, as previously claimed, the $\beta_0$ contribution to $\widehat{\f V}_g^{(2,0)}$ gets cancelled.

The matching for the axial-vector sector and is completely analogous. In the gluon transversity sector it is even simpler because there is no mixing; the quark CF is zero.

We have found the calculation of the convolution integrals in eqs. \eqref{eq: quark two-loop matching} and \eqref{eq: gluon two-loop matching} to be most conveniently calculated using the expressions in eqs. \eqref{eq: vector partonic GPD}, \eqref{eq: axial partonic GPD} \eqref{eq: transversity partonic GPD}. After performing the $x$ integration using the delta functions, the remaining integrals over $\alpha$ and $\beta$ can be readily evaluated using \texttt{HyperInt} \cite{Panzer:2014caa}. For the two-loop CFs one only needs the one-loop evolution kernels. In principle one has to be careful with $\gamma_5$ schemes in this regard, but at one-loop the evolution kernels in Larin's scheme and in \cite{Belitsky:2005qn}, which are in the NDR-scheme, coincide.

\section{Weight reduction of transversity CF}

\label{sec: B}

In this section we discuss the origin of the weight reduction by two, observed for the transversity gluon CF. 
Let us consider the matching for $C_{g,T}$.
In the following we use the notation introduced in Appendix \ref{sec: A}.
The factorization formula dictates that the UV-renormalized and double-counting subtracted $C_{g,T}^{(2)}$ in $d$ dimensions takes the form \footnote{Clearly, the $\epsilon$-pole counting is determined by UV-renormalization and factorization.},
\begin{align}
C_{g,T} &=\frac{\alpha_s}{4\pi} {\m C}_{g,T}^{(1,0)} + \Big ( \frac{\alpha_s}{4\pi} \Big )^2 \Big [\frac{1}\epsilon\left( {\m C}_{g,T}^{(2,-1)}+ {\m C}_{g,T}^{(1,0)}\ast \widehat F_{g/g,T}^{(1,-1)}-\beta_0  {\m C}_{g,T}^{(1,0)}\right) \notag
\\
&\quad + {\m C}_{g,T}^{(1,1)} \ast \widehat{F}_{g/g,T}^{(1,-1)} + O(\epsilon)\Big ] + O(\alpha_s^3), \label{eq:CgT-IRS}
\end{align}
where $\m C_{g,T}$ is the unsubtracted, or ``bare'', coefficient function, i.e.
\begin{align}
\m C_{g,T}(x) = \frac{\widehat{\f T}_g(1/x)}{1-x^2},
\end{align}
where $\f T_g(\xi)$ is the transversity amplitude, defined in eq. \eqref{eq: V, A, T}, with the partonic external states in eq. \eqref{eq: transversity ext states}. Eq. \eqref{eq:CgT-IRS} is symbolical, with an implicit convolution product denoted by $\ast$, see Appendix \ref{sec: A} for details.

The finiteness of $C_{g,T}$ in $\epsilon\to0$ then implies from \eqref{eq:CgT-IRS} that 
\begin{align}
{\mathbf C}_{g,T}^{(2,1)}=- {\mathbf C}_{g,T}^{(1,0)}\ast \widehat F_{g/g,T}^{(1,-1)}+\beta_0  {\mathbf C}_{g,T}^{(1,0)}\, ,
\end{align}
which indicates that the transdentality weight of $ {\mathbf C}_{g,T}^{(2,1)}$ is one from convoluting two weight zero rational functions $ {\mathbf C}_{g,T}^{(1,0)}$ and $\widehat F_{g/g,T}^{(1,-1)}$. On the other hand, the bare two-loop CF ${\mathbf C}_{g,T}^{(n)}$ is generated via a set of master integrals $M_i$ all of which have the following property,
\begin{align}
    M_i(z,\epsilon) = \sum_j \epsilon^j f_{i,j}(z)\, , \qquad\qquad w[f_{i,j+1}(z)] = w[f_{i,j}(z)] + 1
\end{align}
with $w[g(z)]$ being the transcendental weight of function $g(z)$. $ {\mathbf C}_{g,T}^{(n)}$ is then related to the masters (color factors are suppressed),
\begin{align}
     {\mathbf C}_{g,T}^{(2)}(z,\epsilon) = \sum_{i} r_i(z,\epsilon) M_i(z,\epsilon)
\end{align}
through IBP relations hence $r_i(z,\epsilon)$ are rational functions in $z$ and $\epsilon$. This allows us to conclude that ${\mathbf C}_{g,T}^{(2)}(z,\epsilon)$ follow the same weight counting as the masters, i.e.,
\begin{align}
w\left[{\mathbf C}_{g,T}^{(2,m+1)}\right]  = w\left[{\mathbf C}_{g,T}^{(2,m)}\right] + 1\, .
\end{align}
Combined with our previous result $w\left[{\mathbf C}_{g,T}^{(2,-1)}\right]=1$, and $w\left[{\mathbf C}_{g,T}^{(1,0)}*\widehat{F}_{g/g,T}^{(1,-1)}\right] = w\left[{\mathbf C}_{g,T}^{(1,0)}\right] +1 = 2$, we finally conclude, without explicit calculations, that $w\left[C_{g,T}^{(2)}\right]=2$.

It remains to be seen or argued if the patterns 
\begin{align*}
    w\left[C_{g,T}^{(n)}\right] = w\left[C_{i}^{(n-1)}\right]=w\left[C_{i}^{(n)}\right]-2\, ,~~ w\left[C_{i}^{(n)}\right] = w\left[\widetilde C_{i}^{(n)}\right]\, , \quad \text{where} ~~ n\geq3 ~~\text{and}~~ i = q, g
\end{align*} 
persist to higher orders in $\alpha_s$. We leave this question for future work.

%
%
%

\newpage
%

%
%
%




\providecommand{\href}[2]{#2}\begingroup\raggedright\endgroup

\end{document}